\documentclass{IEEEtran}

\usepackage[colorlinks,urlcolor=blue,linkcolor=blue,citecolor=blue]{hyperref}

\usepackage{graphicx}
\usepackage{cite}
\usepackage{amsmath,amssymb,amsfonts}
\usepackage{algorithmic}
\usepackage{graphicx}
\usepackage{textcomp}
\usepackage{xcolor,colortbl}

\begin{document}

\title{Adaptive MIMO Radar Architecture for Energy-Efficient Wireless Sensing in the D-Band}

\author{Subbarao Korlapati and Reza Nikandish, \IEEEmembership{Senior Member, IEEE}
}

\maketitle

\begin{abstract}
\boldmath
The D-band offering an untapped wide bandwidth is promising for high data rate communication and high-resolution wireless sensing. However, these potentials are hindered by the low performance and energy efficiency of the D-band circuits and systems.
We present an adaptive multi-input multi-output (MIMO) radar architecture for energy-efficient wireless sensing in the D-band, leveraging a reconfigurable 2D array of radar transceiver front-ends, a scaling approach for the receiver (RX) signal-to-noise ratio (SNR) and the transmitter (TX) output power ($P_{\rm TX}$) with target distance, and dynamic selection of the direction-of-arrival (DOA) estimation algorithm. 
The reconfigurable radar array, providing an adaptive radar resolution, enhances the energy efficiency by reducing power consumption in the radar RF front-end and lowering the computational complexity in the radar back-end.
The RX SNR and the TX output power are scaled with the distance as ${\rm SNR} \propto d^{-p}$ and $P_{\rm TX} \propto d^{4-p}$, where $0 < p < 4$, leading to more efficient resource allocation in varying target distance conditions. 
Additionally, DOA estimation results using MUSIC and MVDR algorithms indicate that the optimum algorithm, in terms of the accuracy and computational complexity, should be selected based on the number of radar array elements.
Furthermore, we develop a hardware model for the MIMO radar RF front-end to evaluate the power consumption of the TX, RX, and local oscillator (LO) distribution network. It is shown that the power consumption of the LO distribution network, which can dominate the power consumption for a large MIMO radar, can be minimized through a distribution strategy for LO amplifiers employed for compensating passive losses. Performance of the adaptive MIMO radar is evaluated in the free-space and the through-wall indoor sensing scenarios in the D-band.
\end{abstract}

\begin{IEEEkeywords}
Adaptive, D-band, millimeter-wave, multi-input multi-output (MIMO), radar, receiver, system-on-chip (SoC), transmitter, wireless sensing.
\end{IEEEkeywords}

\section{Introduction}
\label{sec:introduction}
\IEEEPARstart{M}{illimeter-Wave} frequency bands above 100\,GHz are promising for the next generation of wireless communication, sensing, and imaging systems. The extremely wide bandwidths available in these bands can enable data rates of multi 100\,Gbps for communication and millimeter-scale resolutions for sensing and imaging. The D-band (110--170\,GHz) is a candidate for 6G communications \cite{6g-white-2020, Rappaport-access2019, maiwald-proc2022}.

Wireless sensing using radars is a key technology for emerging applications such as integrated radar-communication systems, autonomous vehicles, gesture recognition for human-computer interaction, contactless health monitoring, and medical imaging. Recently, several developments are presented towards the realization of wireless communications and sensing above 100\,GHz. These developments, in the system and circuit levels, include experimental verification of the free-space path loss (FSPL) model \cite{Rappaport-access2019}, characterization of the propagation channel \cite{kim-tap2015}, modeling and measurement of surface scattering \cite{Rappaport-access2019}, loss measurements of outdoor and indoor materials \cite{Rappaport-access2019, beelde-access2021}, and implementation of system-on-chip (SoC) transceivers \cite{maiwald-proc2022, Visweswaran-jssc2021, sense-lmwc-2022, Furqan-fullyIntgrated-2019, Ahmad-W-Band-D-Band, Deng-D-Band-2023, Zandieh-155GHz-FMCW, Kucharski-Multi-Purpose-2017, D-Band-Low-Power-2022}. The state-of-the-art D-band SoC radars have achieved a bandwidth of 30\,GHz, chirp sweep time of 1--100\,$\rm{\mu s}$ and the slope of 1--8 GHz/$\rm{\mu s}$, the transmitter (TX) output power 10--13\,dBm (10--20\,mW) and the receiver (RX) noise figure (NF) of 8--12\,dB \cite{Visweswaran-jssc2021, sense-lmwc-2022, Furqan-fullyIntgrated-2019, Ahmad-W-Band-D-Band, Deng-D-Band-2023, Zandieh-155GHz-FMCW, Kucharski-Multi-Purpose-2017, D-Band-Low-Power-2022}. These systems have been implemented using Silicon processes to achieve high integration levels. However, the lower output power, efficiency, and noise performance of Silicon compared to III-V processes (e.g., GaN \cite{nikandish-jmw2024}) lead to very low \textit{energy efficiencies}. This is a significant challenge for the D-band wireless sensing, where a large array of radars is required to achieve high resolutions for detecting and tracking the targets. 

A multi-input multi-output (MIMO) radar is usually used in practical applications to provide sufficient angular resolutions, increase the spatial diversity, and improve effective SNR of the received signal \cite{fishler-radar2004, li-2007, serio-2020, zhuge-tap2012, peng-tmtt2018, tan-tap2017}. A major issue in conventional MIMO radars is the use of \textit{fixed hardware and algorithm} for directional-of-arrival (DOA) estimation of the targets \cite{maiwald-proc2022, Puglielli-proc2016, Torkildson-twc2011, nikandish-tradar2023, arnold-tap2019}. This leads to inefficient allocation of the radar resources in varying operational conditions.  

We propose an adaptive MIMO radar architecture with the following key features: 1) the effective number of array elements is reconfigured based on the required accuracy of the DOA estimation to save the power consumption and reduce the computational complexity, 2) the output power of the TX and the required SNR of the RX are scaled with the distance to reduce the power consumption, and 3) the DOA estimation algorithm is selected based on the number of array elements to enhance accuracy and reduce the computational complexity. We develop a hardware model for the MIMO radar to estimate the power consumption of the TX, RX, and local oscillator (LO) distribution network. Performance of the adaptive MIMO radar is evaluated in the free-space and through-wall indoor wireless sensing scenarios, using two popular DOA estimation algorithms, multiple signal classification (MUSIC) \cite{music-paper} and minimum variance distortionless response (MVDR) \cite{mvdr-paper}. 

The paper is structured as follows. In Section II, the adaptive MIMO radar architecture is presented. In Section III, the MIMO radar front-end is discussed and a circuit model is developed for the radar system to estimate the power consumption. In Section IV, the MIMO radar back-end including the DOA estimation algorithms is discussed. In section V, performance of the adaptive MIMO radar is evaluated for D-band wireless sensing. Finally, the concluding remarks are presented in Section VI.

\section{Adaptive MIMO Radar Architecture}

\subsection{Adaptive Number of Array Elements}
\begin{figure}[!t]
    \centering
    \includegraphics[width =  \columnwidth]{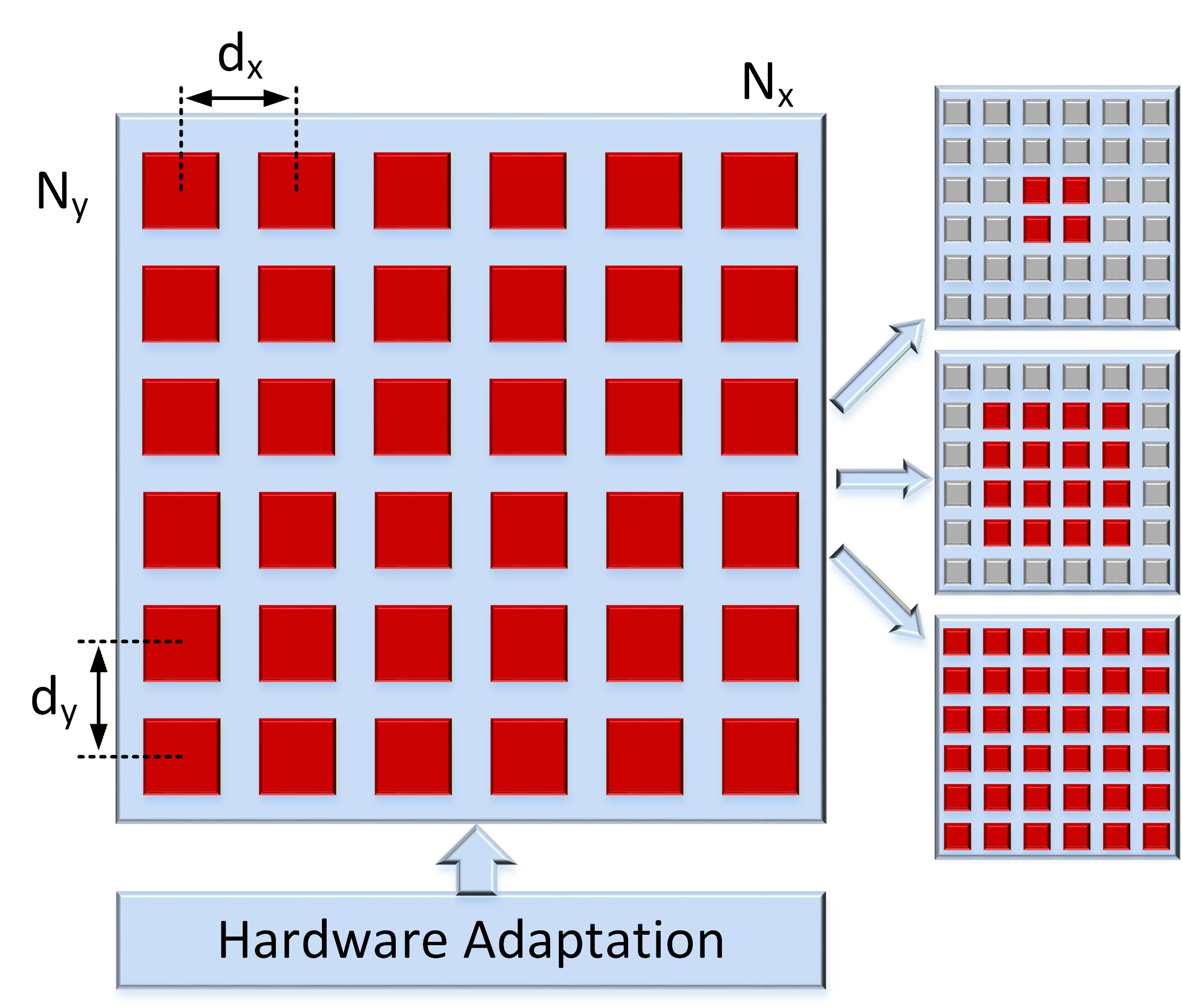}
    \caption{Architecture of the 2D MIMO radar comprising $N_x \times N_y$ virtual arrays. The effective array size can be adaptively reconfigured by disabling certain TX and RX elements to improve the energy efficiency.}
    \label{fig:adaptive_mimo}
\end{figure}

We consider a uniform 2D array of $N_x\times N_y$ radar elements, as shown in Fig. \ref{fig:adaptive_mimo}. This array can be constructed as a virtual array comprising $N_{\rm {TX}}$ transmitters and $N_{\rm {RX}}$ receivers. The number of virtual array elements is given by $N_xN_y = N_{\rm {TX}} N_{\rm {RX}}$. The distances are selected such that the inter-element spacing is $\lambda/2$. In the D-band, $\lambda/2 \approx {\rm {1\,mm}}$, which allows large arrays to be integrated on a small physical area, for example, a 100-element array on a ${\rm {1\,cm \times 1\,cm}}$ area. This allows the MIMO radar to be integrated into electronic devices for wireless sensing capabilities. 

The effective size of the radar array can be reconfigured by disabling certain TX and RX elements, as shown in Fig. \ref{fig:adaptive_mimo}. This decreases the angular resolution of the radar but also reduces the power consumption of RF circuits and the computational complexity of the digital signal processing (DSP) system. The key idea is to adaptively change the array size based on the required resolution in the operational conditions to increase energy efficiency of the sensing system.

\subsection{Adaptive Radar Resolution}
A FMCW MIMO radar can detect 3D coordinates of the targets. The FMCW radar elements transmit a frame of multiple chirp signals, correlates the received target reflection with the transmit signal, and detects the target range using the signal time of flight. The angular information of the targets are extracted using the signals transmitted and received by the 2D MIMO array. The range resolution is dependent on the chirp signal bandwidth ($\Delta R = 2c/B$), while the angular resolution can be adaptively controlled by the active number of array elements, transmitted power, and received SNR. 

The array factor for a 2D array can be derived as
\begin{equation}
    \label{AF}
    {AF}(\theta, \phi) = \sum_{n=1}^{N_y} \sum_{m=1}^{N_x} e^{j[(m-1) \psi_x + (n-1)\psi_y]}
\end{equation}
\begin{equation}
    \label{psi_x}
    \psi_x = \beta d_x\sin{\theta} \cos{\phi}
\end{equation}
\begin{equation}
    \label{psi_y}
    \psi_y = \beta d_y\sin{\theta} \sin{\phi}
\end{equation}
where $\beta = {2\pi}/{\lambda}$ is the wave number, $d_x$ and $d_y$ are the inter-element spacing in the $x$ and $y$ directions, and $(\theta, \phi)$ are the elevation and azimuth angles \cite{Stutzman-2012}. For a uniform array, $d_x = d_y = \lambda/2$. 
A larger array provides higher directivity, narrower beamwidth, and lower side lobe level. 
The angular resolution of the MIMO radar can be derived as the separation between two consecutive nulls in $AF(\theta, \phi)$ versus the azimuth or elevation angle. The antenna elements have a limited beamwidth, which can be modeled as $|\phi| < \phi_{m}$ and $|\theta| < \theta_{m}$. This leads to the angular resolutions as
\begin{equation}
    \label{azimuth_reolution_max}
     \Delta \phi \approx \frac{2}{N_x \cos{\phi_{m}}}
\end{equation}
\begin{equation}
    \label{elevation_reolution_max}
     \Delta \theta \approx \frac{2}{N_y \cos{\theta_{m}}}.
\end{equation}
The number of array elements should be selected based on the required angular resolutions and the beamwidth of the antennas. In the adaptive MIMO radar with varying number of array elements based on the operating conditions, this helps to reduce the power consumption and the computational complexity.

In the presence of noise and interference signals, certain signal processing algorithms such as MUSIC should be used to extract the signal features and estimate the DOA. Effective SNR of the MIMO radar under optimal signal combining is improved by a factor equal to the number of virtual array elements compared to SNR of each radar element
\begin{equation}
    \label{snr_mimo}
     {\rm {SNR_{MIMO}}} = {\rm {SNR_{SISO}}} + 10\log_{10}(N_{\rm {TX}} N_{\rm {RX}}). 
\end{equation}
For example, by doubling the number of TX or RX elements, the SNR of the MIMO radar is improved by 3\,dB. The accuracy of the DOA estimation is dependent on the number of array elements, the SNR of the received signal, and the algorithm. Therefore, an adaptive radar resolution can be realized by varying the number of array elements.

\subsection{Scaling of SNR and Output Power with Distance}
In free space, using the fundamental radar range equation, the SNR of the received signal can be derived as
\begin{equation}
    \label{radar_snr}
     {\rm {SNR}} = \frac{P_{TX} G_{TX} G_{RX} \lambda^2 \sigma T_{meas}}{(4\pi)^3 kTF d^4},
\end{equation}
where $P_{TX}$ is the TX output power, $G_{TX}$ and $G_{RX}$ are gain of the TX and RX antennas, $\lambda$ is the signal wavelength, $\sigma$ is the radar cross section (RCS) of the target, $T_{meas}$ is the radar measurement time, $k = \rm {1.38 \times 10^{-23}\,J/K}$, $T$ is the absolute temperature, $F$ is the RX noise factor, and $d$ is the target distance.

In classic DOA estimation approaches, the SNR is considered as a parameter independent of the target distance. However, this approach requires scaling of the TX output power with the distance as $P_{\rm {TX}} \propto d^4$ [see (\ref{radar_snr})]. This can lead to an impractical TX output power requirement at long distances. On the other hand, if we assume a fixed TX output power, the RX SNR will decrease with the distance as ${\rm {SNR}} \propto d^{-4}$. A compromise between the two extremes is to assume that both the TX output power and the RX SNR are scaled with the distance. 

\begin{figure}[!t]
  \begin{center}
  \includegraphics[width=\columnwidth]{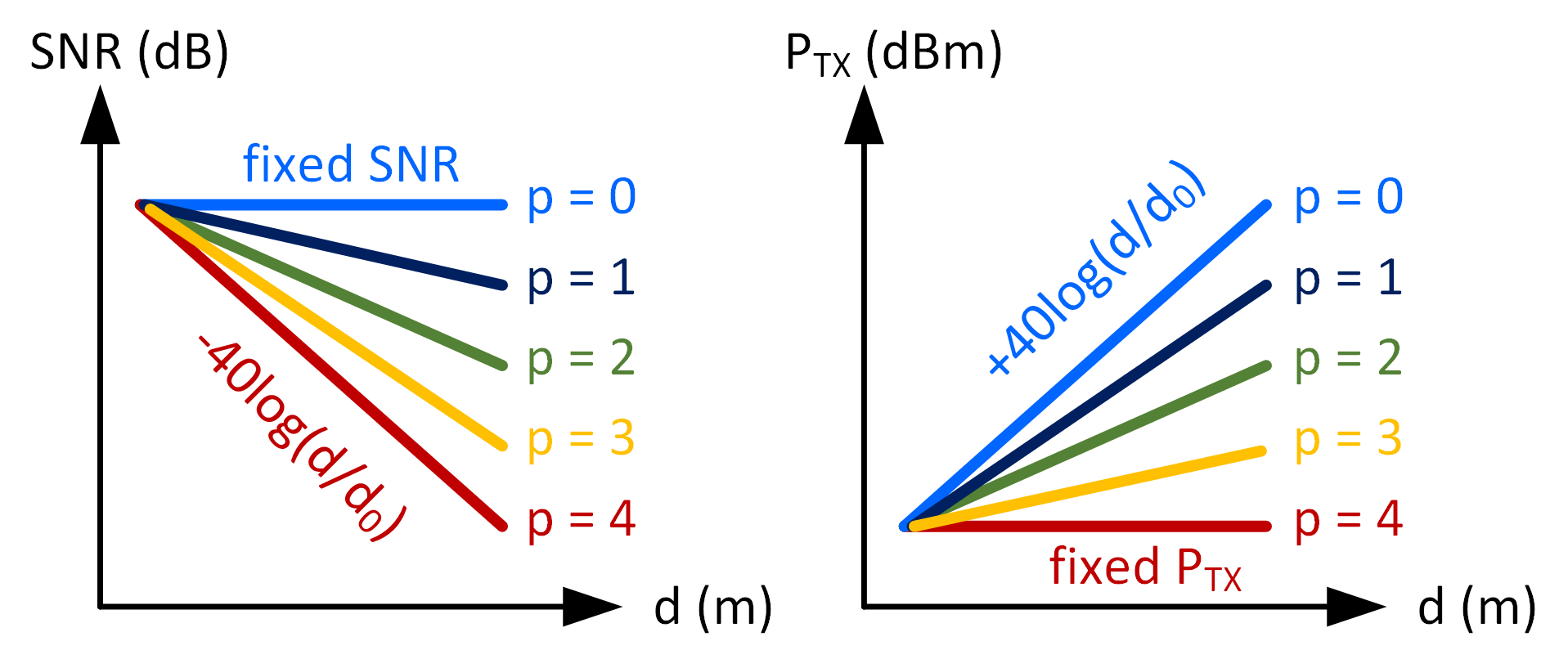}
  \caption{Scaling of the RX SNR and the TX output power with distance. In the fixed SNR scenario (blue), the TX power should be increased with the slope of $+40\log(d/d_0)$, while in the fixed TX power scenario (red), the SNR is reduced with the slope of $-40\log(d/d_0)$. A compromise can be achieved by selecting a scaling exponent in the range $0<p<4$.}
  \label{fig:SNR_PTX_d}
  \end{center}
\end{figure}

We propose a scaling approach for the RX SNR and the TX output power with the distance as 
\begin{equation}
    \label{snr_scaling}
     {\rm {SNR}} (d) =  {\rm {SNR}} (d_0) \left( \frac{d_0}{d} \right)^p
\end{equation}
\begin{equation}
    \label{PTX_scaling}
     P_{\rm {TX}} (d) =  P_{\rm TX} (d_0) \left( \frac{d_0}{d} \right)^{p-4},
\end{equation}
where $p$ is the scaling exponent that should be selected in the range $0 < p < 4$ and $d_0$ is the reference distance. The TX output power is set to achieve a certain RX SNR at a reference distance $d_0$, and it is scaled according to (\ref{PTX_scaling}) at other distances. The classic constant SNR scenario is achieved for $p=0$, while the case that the TX output power is constant corresponds to $p=4$. In Fig. \ref{fig:SNR_PTX_d}, the profiles of RX SNR and TX output power versus distance are shown for values of $p$ using (\ref{snr_scaling}) and (\ref{PTX_scaling}). We select $p=2$ for a compromise between the slopes of the SNR and the required TX output power.

\subsection{Adaptive MIMO Waveforms}
\begin{figure}[!t]
    \centering
    \includegraphics[width = 0.9 \columnwidth]{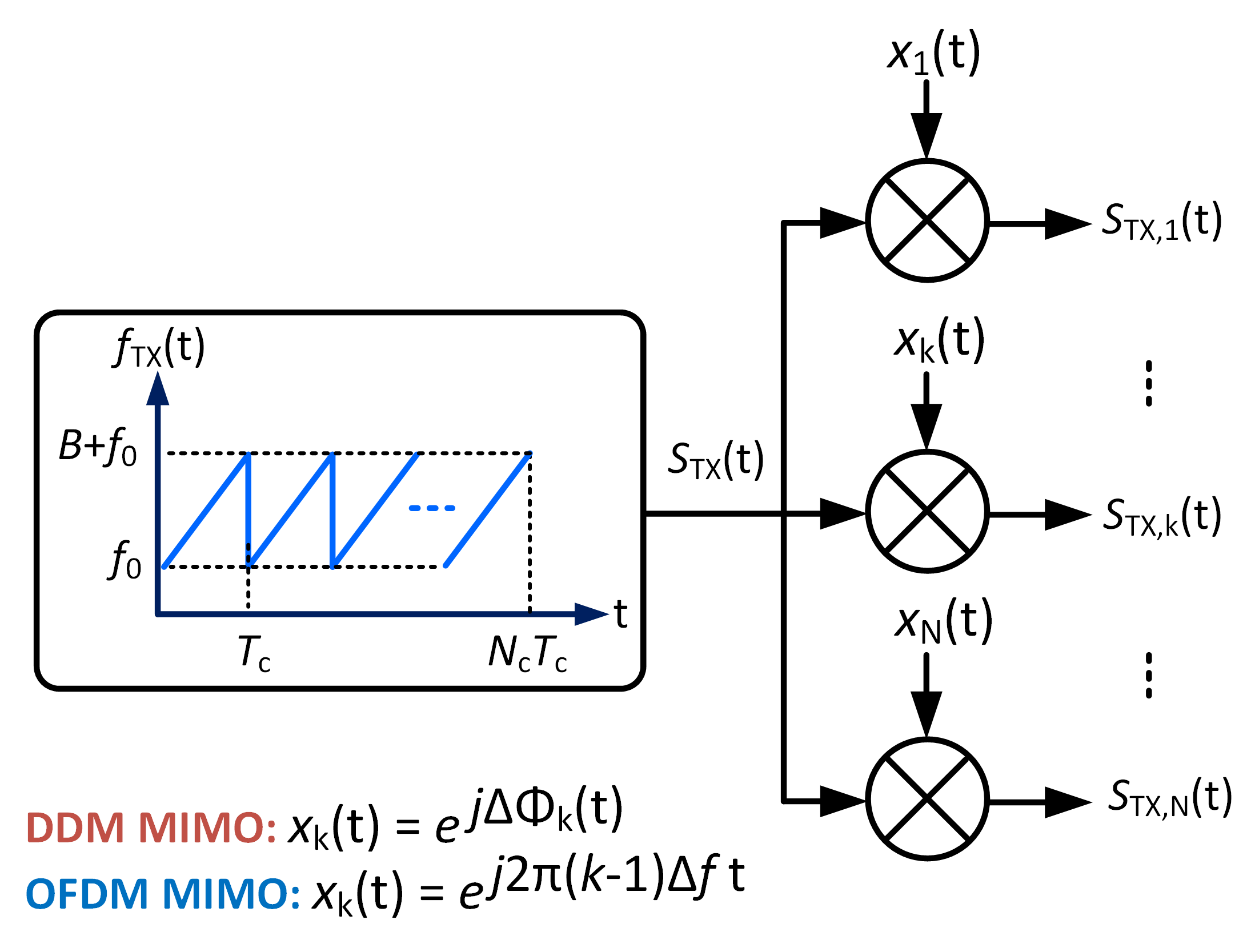}
    \caption{Conceptual orthogonal waveform generation for MIMO FMCW radar. The chirp signal $S_{\rm {TX}}(t)$ is modulated by the sequence $x_k(t)$ to realize the transmit signals $S_{\rm TX,k}(t)$ for $k = 1, ..., N_{\rm {TX}}$. The MIMO waveforms are adaptively varied as the number of TX elements change.}
    \label{fig:mimo_waveform}
\end{figure}

In the MIMO radar, orthogonal waveforms are transmitted to simplify the separation of received signals \cite{sun-spm-2020}. This can be achieved by applying a multiplexing scheme to modulate the chirp sequence signal $S_{\rm {TX}}(t)$ into the TX signals, $S_{\rm {TX,k}}(t)$ for $k = 1, ...,N_{\rm {TX}}$, as shown in Fig. \ref{fig:mimo_waveform}. Time-division multiplexing (TDM) is the simplest scheme for MIMO radar where only one TX transmits at a time slot
\begin{equation}
    \label{x_tdm}
     x_{k,{\rm TDM}}(t) = p \left(\frac{t - \Delta t_k}{T} \right),
\end{equation}
where $p(t)$ is the pulse function. The time offset is selected as $\Delta t_k = (k-1)T/N_{TX}$, which should be adaptively varied based on the number of active TX elements. However, the TDM scheme degrades the overall transmitted power, which is crucial for mm-wave bands. 

In the Doppler-division multiplexing (DDM) scheme, a phase code is applied with a certain pattern to realize the TX signals
\begin{equation}
    \label{x_ddm}
     x_{k,{\rm DDM}}(t) = e^{-j \Delta \Phi_k(t)},
\end{equation}
where the phase offset can have various forms. A simple approach is to select it as $\Delta \Phi_k(t) = (k-1)2\pi/N_{TX}$, which should be adaptively varied based on the number of active TX elements.

In the orthogonal frequency-division multiplexing (OFDM) scheme, the TX signal is modulated by different carrier frequencies, which are usually shifted by an offset frequency
\begin{equation}
    \label{x_ofdm}
     x_{k,{\rm OFDM}}(t) = e^{-j [2\pi (k-1) \Delta f] t}.
\end{equation}
The offset frequency should be adjusted based on the number of active TX elements if a certain bandwidth is considered for the OFDM scheme, $\Delta f = B_{\rm OFDM}/N_{TX}$. 

In the DDM and OFDM MIMO schemes, all of the radar transmitters operate simultaneously and the total output power is enhanced. Therefore, we assume that the MIMO radar is realized using DDM or OFDM schemes with \textit{adaptive waveforms} varied by the number of active TX elements.

\section{MIMO Radar Front-End}

\subsection{Front-End Architecture}

\begin{figure}[!t]
    \centering
    \includegraphics[width = 0.9 \columnwidth]{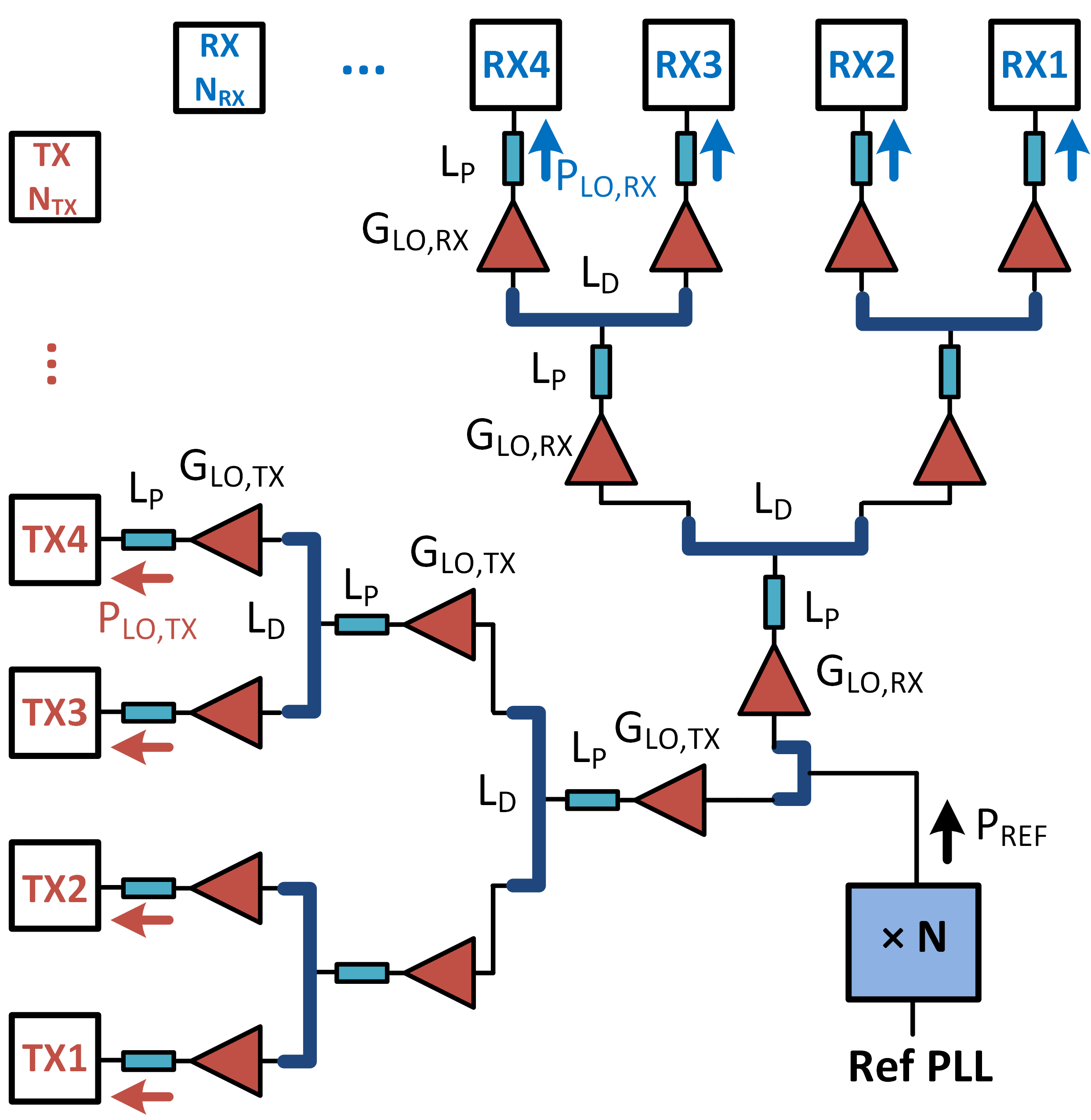}
    \caption{Front-end architecture of the 2D MIMO radar realized using two linear arrays of the TX and RX elements. The LO signal distribution network comprises the losses of the LO signal paths $L_P$, the losses of the power dividers $L_D$, and the LO amplifiers with power gain of $G_{\rm LO,TX}$ and $G_{\rm LO,RX}$.}
    \label{fig:physical_array}
\end{figure}

 We consider the MIMO radar front-end shown in Fig. \ref{fig:physical_array}, which comprises two uniform linear arrays of the RX and TX elements. This is equivalent to a 2D virtual array of $N_{\rm TX} N_{\rm RX}$ elements. This architecture enables a symmetric LO distribution to the TX and RX elements. This leads to \textit{correlated phase noise} of the TX and RX signals, which in a FMCW radar suppresses the phase noise in the received IF signal \cite{Visweswaran-jssc2021}. The LO distribution network is a critical component of a MIMO radar, especially in the D-band, and can dominate the power consumption of the overall radar system. Therefore, it is imperative to minimize the associated losses and power consumption. We use this architecture to estimate the losses and power consumption of the LO distribution network, the required TX output power, and the power consumption of the MIMO radar front-end. 

The D-band antenna array can be implemented using the antenna-in-package (AiP) technology and directly flip-chip attached to the chip die \cite{gu-jmw2021}. This allows the integration of large MIMO radars with high radiation efficiencies (e.g., over 80\%). 
 
\subsection{LO Signal Distribution}

The LO generation comprises a phase-locked loop (PLL) and frequency multiplier(s) with low spurs, wide bandwidth, good chirp linearity, and high DC-to-RF efficiency. The reference signal can be generated by an external PLL. The LO distribution network, as shown in Fig. \ref{fig:physical_array}, comprises the routing paths to the TX and RX elements, and the LO amplifiers to compensate for the losses and maintain the signal integrity. 

Frequency multipliers can be used at different locations in the MIMO radar to upconvert the low-frequency reference signal to the D-band LO signals.
We can envision two extreme cases for the deployment of frequency multipliers. In the first approach, the reference signal is unconverted using a single high-performance frequency multiplier (usually comprising multiple stages) to the final RF signals and is then distributed between the TX and RX systems. The drawback is the higher loss of the LO distribution network and the higher power consumption of the LO amplifiers. In the second approach, a low-frequency reference signal is distributed between the TX and RX paths, and is upconverted by dedicated frequency multipliers in the all TX and RX systems. This can reduce the losses of the LO distribution network and the power consumption of the LO amplifiers. However, this needs a total of $N_{TX} + N_{RX}$ frequency multipliers which can lead to high power consumption for large MIMO radars. The optimal architecture should be between these two extremes, but it depends on many factors on the circuit and process level. In this work, we use the first approach to perform system simulations. 

\begin{figure}[!t]
    \centering
    \includegraphics[width = \columnwidth]{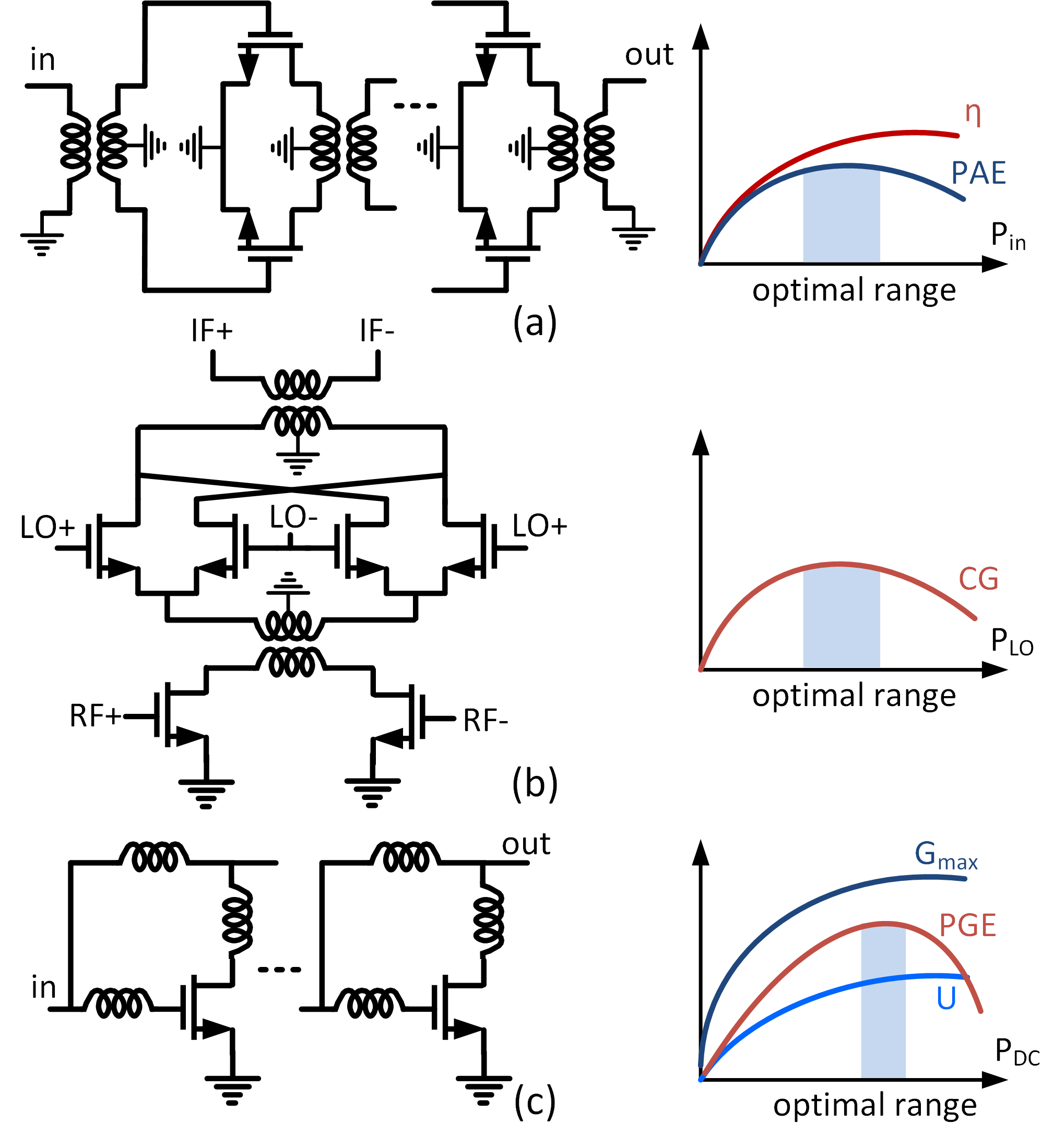}
    \caption{Structure of typical mm-wave circuits used in the radar: (a) power amplifier circuit, and its efficiency $\eta$ and power-added efficiency (PAE) versus input power, (b) mixer circuit and its conversion gain (CG) versus LO power, (c) amplifier circuit with power gain boosting, and the maximum power gain $G_{\rm max}$, unilateral power gain $U$, and the power gain efficiency (PGE), $ {\alpha = G_P/{P_{DC}}}$, for an amplifier versus power consumption.}
    \label{fig:circuits}
\end{figure}

In Fig. \ref{fig:physical_array}, the output power of the reference signal after being upconverted by the frequency multiplier is $P_{\rm REF}$, and after power division between the routing paths it reaches to $P_{\rm LO,TX}$ in the TX and $P_{\rm LO,RX}$ in the RX systems. Both of these LO power levels should be within certain ranges for their efficient operation. In the TX, the LO power should be high enough to push the PA into saturation and achieve high efficiency, as shown in Fig. \ref{fig:circuits}(a). However, an excessively high LO power can degrade gain, power-added efficiency (PAE), defined as $\rm{PAE = (P_{out} - P_{in})/P_{DC}}$, and the linearity of the PA. In the RX, the LO power should be high enough to enable switching of the transistors operating as the mixer, as shown in Fig. \ref{fig:circuits}(b). On the other hand, for excessively large LO power levels, the conversion gain of the mixer, defined as $\rm{G_c = P_{out,IF}/P_{in,RF}}$, will degrade due to the operation of the transistor pair in the triode mode during most of the signal cycle \cite{Deng-D-Band-2023}. Therefore, the LO power delivered to TX and RX should be maintained within optimal power ranges,
\begin{equation}
    \label{P_LO}
    P_{\rm LO,min} \leq P_{\rm LO} \leq P_{\rm LO,max},
\end{equation}
where the lower and upper limits of the LO power are dependent on the properties of the circuit and fabrication process.

The reference signal is highly attenuated by the divisions between the routing paths and the losses of on-chip transmissions lines in the D-band. Therefore, the LO amplifiers are distributed along the routing paths to prevent the need for a single high-gain LO amplifier after the reference signal generator. We propose to use \textit{one LO amplifier in each path} with a power gain of $G_{\rm LO,TX}$ in the TX paths and $G_{\rm LO,RX}$ in the RX paths (see Fig. \ref{fig:physical_array}). The loss of the LO paths is dependent on their physical length, layout structure, and properties of the chip fabrication process. We assume that the length of all branches is $\Delta l$ to simplify derivations. This length is dependent on the routing layout, but it can be fairly approximated by a fraction of the signal wavelength, e.g.,, $\Delta l \approx \lambda /2$. The loss per unit length of the LO distribution paths can be estimated using the loss properties of transmission lines in the CMOS process (typically 1--3\,dB/mm in the D-band). The loss of each path with the length of $\Delta l$ is derived as $L_P$.

The LO power delivered to \textit{each} of the TX and RX systems can be derived using Fig. \ref{fig:physical_array} as
\begin{equation}
    \label{LO_TX_Power}
    {P_{\rm LO,TX}} = \frac{L_D(G_{\rm LO,TX} L_P)^{\log_2{N_{TX}}}}{N_{TX}} P_{REF}
\end{equation}
\begin{equation}
    \label{LO_RX_Power}
    {P_{\rm LO,RX}} = \frac{L_D(G_{\rm LO,RX} L_P)^{\log_2{N_{RX}}}}{N_{RX}} P_{REF},
\end{equation}
where $G_{\rm LO,TX}$ and $G_{\rm LO,RX}$ are the gain of LO amplifiers in the TX and RX paths, $L_D$ is the loss of the power dividers, and $L_P$ is the loss of each LO path. These equations can be solved to derive the required gain of the LO amplifiers as 
\begin{equation}
    \label{LO_TX_Gain}
    G_{\rm LO,TX} = \frac{1}{L_P} \left( \frac{N_{TX}}{L_D} \frac{{P_{ LO,TX}}}{P_{REF}}  \right)^{\frac{1}{\log_2{N_{TX}}}}
\end{equation}
\begin{equation}
    \label{LO_RX_Gain}
    G_{\rm LO,RX} = \frac{1}{L_P} \left( \frac{N_{RX}}{L_D} \frac{{P_{ LO,RX}}}{P_{REF}}  \right)^{\frac{1}{\log_2{N_{RX}}}}.
    \end{equation}
The gain of the LO amplifiers can be adjusted to compensate for the losses or amplify the reference signal if required.

The total number of the LO amplifiers in the TX path can be calculated using Fig. \ref{fig:physical_array} as
\begin{equation}
    \label{LO_TX_amplifiers_1}
    N_{\rm LO, TX} =  1 + 2 + ... + \frac{1}{2}N_{TX} + N_{TX} = \sum_{k=0}^{\log_2{N_{TX}}} 2^k,
\end{equation}
which using $\sum_{k=0}^{n}a^k=(a^{n+1}-1)/(a-1)$ is derived as
\begin{equation}
    \label{LO_TX_amplifiers}
    N_{\rm LO, TX} = 2 N_{TX} -1.
\end{equation}
Similarly, in the RX path, the number of the LO amplifiers can be derived as
\begin{equation}
    \label{LO_RX_amplifiers}
    N_{\rm LO, RX} = 2 N_{RX} -1.
\end{equation}
These results will be used to estimate the power consumption of the LO distribution network in Section III-D. 

\subsection{Power Consumption of Radar Front-End}

The power consumption of the MIMO radar is dependent on the number of array elements, system architecture of the radar TX and RX, LO distribution network, circuit structures, and fabrication process. In the D-band, the power consumption of the RX circuits is comparable with that of the TX circuits, and the LO distribution network can have a significant contribution which dominates the total power consumption. The power consumption of the radar front-end can be estimated as
\begin{equation}
    \label{PDC}
    P_{\rm {DC}} \approx P_{\rm {DC, TX}} + P_{\rm {DC, RX}} + P_{\rm {DC, LO}}.
\end{equation}
We derive each of the three terms versus the number of array elements and properties of the TX, RX, and the LO distribution circuits.

\subsubsection{TX Power Consumption}

The required TX output power for each TX element of the array can be derived using (\ref{radar_snr}), (\ref{snr_scaling}), and (\ref{PTX_scaling}) as
\begin{equation}
    \label{radar_ptx}
     P_{\rm TX}(d)= \frac{(4\pi)^3 kTF d_0^p {\, \rm {SNR}}(d_0)}{G_{TX} G_{RX} \lambda^2 \sigma T_{meas}} d^{4-p}.
\end{equation}
This indicates that for $p<4$, the required TX output power increases with the distance (see Fig. \ref{fig:SNR_PTX_d}). Conventional DOA estimation approaches assume that the SNR is independent of the target distance, that is $p=0$. However, (\ref{radar_ptx}) indicates that this requires the TX power to be increased proportional with $d^4$. This can result in high power consumption in long indoor distances. In our proposed approach, we scale both the SNR and the TX output power with distance by selecting $p=2$. This can significantly reduce the required TX output power and the power consumption of the MIMO radar. 

In the TX system, the power consumption of each element is given by $P_{\rm TX}/\eta_{\rm TX}$, where $\eta_{\rm TX}$ is the TX efficiency. The power consumption of the all TX elements can be derived as
\begin{equation}
    \label{TX_PDC}
    P_{\rm {DC, TX}}(d) \approx \frac{N_{\rm {TX}} P_{\rm {TX}}(d)}{\eta_{\rm {TX}}}.
\end{equation}
In the MIMO radar architecture considered in this work, the TX efficiency can be approximated by the PA efficiency, which is in the order of 5--10\% for silicon-based PAs in the D-band \cite{Visweswaran-jssc2021, Furqan-fullyIntgrated-2019, Deng-D-Band-2023, Zandieh-155GHz-FMCW, D-Band-Low-Power-2022, tang-tcas2020}. It is assumed that the TX circuits are reconfigured as the TX output power is scaled with the distance so that the efficiency remains close to the PA peak efficiency. For the FMCW radars where a saturated PA can be used, this is less challenging than the communication systems using the amplitude-modulated signals. As a result, the power consumption of the TX system also scales with the distance.

\subsubsection{RX Power Consumption}
The RX circuits with major power consumption include the low-noise amplifier (LNA), mixer, and baseband circuits (amplifiers and filters),
\begin{equation}
    \label{RX_PDC}
    P_{\rm {DC, RX}} \approx N_{\rm {RX}} ( P_{\rm {DC, LNA}} + P_{\rm {DC, MIX}} + P_{\rm {DC, BB}}).
\end{equation}
 The power consumption of these circuits is dependent on many factors on the circuit and process level and, as a result, cannot be accurately predicted before the circuits are designed. We can assume a fixed power consumption for each RX and scale it with the number of RX elements in the array. To find a practical estimate, we investigated the power consumption of the TX, RX, and LO parts in the D-band radars presented in the literature and noted that the power consumption of each RX element is usually on the order of 0.5x--1x the power consumption of each TX element \cite{Visweswaran-jssc2021, Ahmad-W-Band-D-Band, Deng-D-Band-2023, Kucharski-Multi-Purpose-2017}. Therefore, we estimate the power consumption of an RX element using that of an TX element as $P_{\rm {DC, RX}}/N_{\rm {RX}} \approx k P_{\rm {DC, TX}}/N_{\rm {TX}}$ with $k \approx 0.75$. We should note that the RX power consumption is not scaled with the TX output power (unlike the TX). Therefore, $P_{\rm {DC, TX}}$ should be evaluated using (\ref{TX_PDC}) at a fixed target distance, such as the reference distance $d_0$.

\subsubsection{LO Power Consumption}
The LO distribution network comprises the LO amplifiers and frequency multiplier(s). The LO amplifiers are designed to achieve the maximum power gain, which is different with the design criteria of the PA and LNA in the TX and RX elements. We can define an efficiency metric namely the \textit{power gain efficiency} for an LO amplifier as the ratio of the power gain to the power consumption
\begin{equation}
    \label{LO_eff}
    \alpha = \frac{G_{P}}{P_{DC}}.
\end{equation}
In mm-wave bands above 100\,GHz, the gain of transistors is very low and usually an embedding passive network is used to boost it \cite{Khatibi-tmtt2018, park-jssc2019}. The maximum power gain of the transistor is given by
\begin{equation}
    \label{G_max}
    G_{\rm max} = ( \sqrt{U} + \sqrt{U-1} )^2,
\end{equation}
where $U$ is the unilateral power gain (Mason’s invariant), dependent on the size and bias of the transistor and the operating frequency \cite{Khatibi-tmtt2018}. By increasing the bias current, $U$ initially increases but after a certain current level it is saturated [see Fig. \ref{fig:circuits}(c)]. Therefore, we can assume that the transistor is biased at an optimum current with an associated unilateral power gain of $U_{opt}$. Using (\ref{LO_eff}) and (\ref{G_max}), we can conclude that this optimal current leads to the maximum $\alpha$ [see Fig. \ref{fig:circuits}(c)]. Usually, multiple amplifier stages should be cascaded to achieve the desired power gain. 

The power consumption of the all LO amplifiers can be derived using (\ref{LO_TX_amplifiers}), (\ref{LO_RX_amplifiers}), and (\ref{LO_eff}) as 
\begin{equation}
    \label{PDC_LO}
    P_{\rm DC,LO} = P_{M} + \frac{(2 N_{TX} -1) G_{\rm LO,TX}}{\alpha_{\rm LO,TX}}+ \frac{(2 N_{RX} -1) G_{\rm LO,RX}}{\alpha_{\rm LO,RX}},
\end{equation}
where $P_{M}$ is the power consumption of the frequency multiplier(s). This equation along with (\ref{LO_TX_Gain}) and (\ref{LO_RX_Gain}) indicate that the power consumption of the LO network is dependent on the number of TX and RX elements, the power gain from the reference to the LO signals, that is $P_{\rm TX,LO}/P_{\rm REF}$ and $P_{\rm RX,LO}/P_{\rm REF}$, the losses of the passive components, $L_P$ and $L_D$, and the power gain efficiency of the transistors, $\alpha_{\rm LO,TX}$ and $\alpha_{\rm LO,RX}$. As we have considered a front-end architecture with a single frequency multiplier, $P_M$ is not scaled with the number of TX and RX elements.

\subsubsection{System Simulations}

\begin{figure}[!t]
    \centering
    \includegraphics[width = 0.95 \columnwidth]{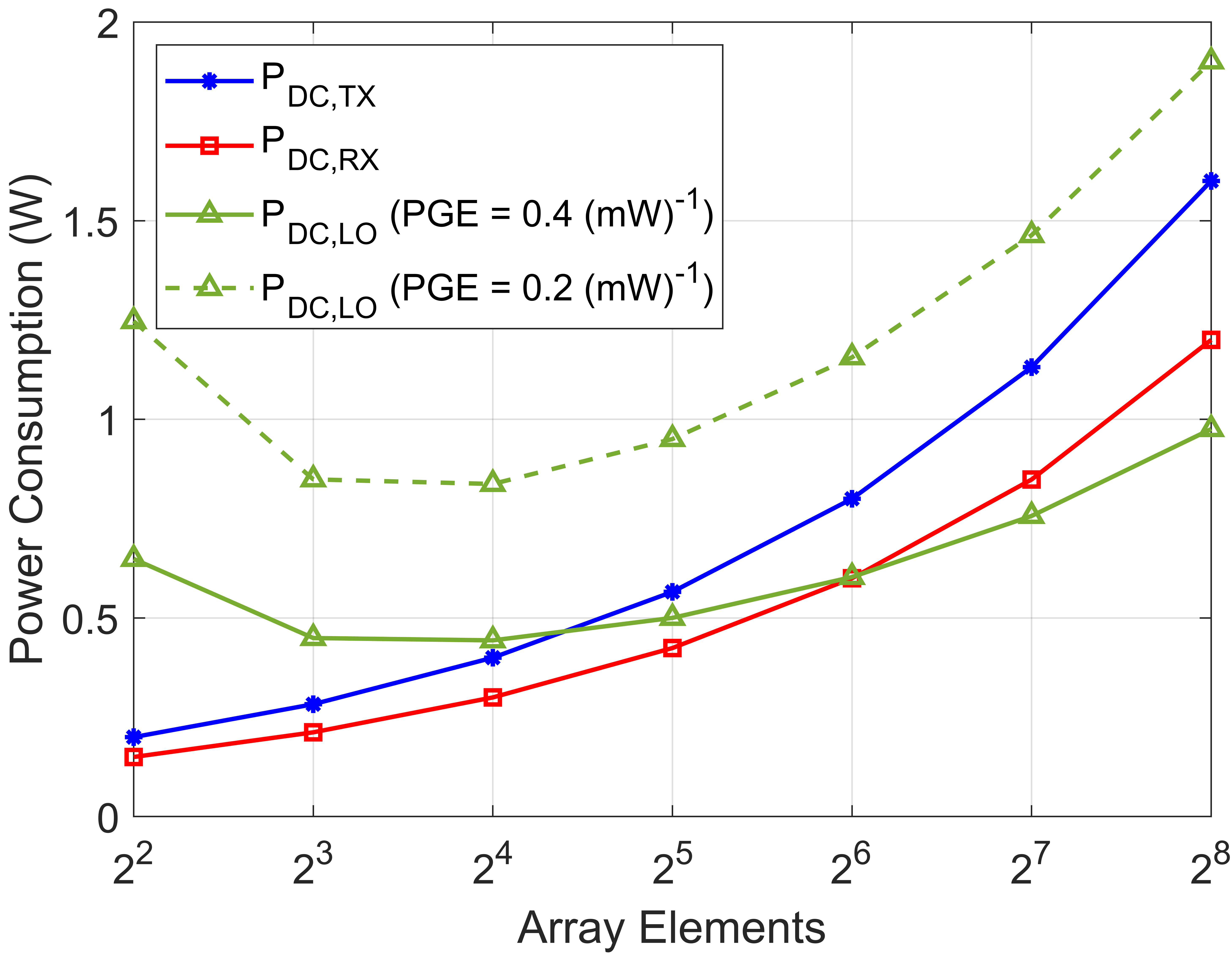}
    \caption{Power consumption of the TX, RX, and LO distribution network versus the number of MIMO radar elements. The LO power consumption is shown for two values of the power gain efficiency (PGE) $\alpha$. The power consumption of the radar front-end is dominated by the LO network for small arrays, while for large arrays it is dominated by the TX when $\alpha$ is high, and by the LO network if $\alpha$ is low.}
    \label{fig:PDC_TX_RX_LO}
\end{figure}

The power consumption of the TX, RX, and LO distribution network versus the number of MIMO radar elements $N_{\rm TX}N_{\rm RX}$ is shown in Fig. \ref{fig:PDC_TX_RX_LO}. The following assumptions are applied. The TX output power and efficiency are 10\,dBm and 10\%, the target distance is 1\,m, and the power consumption of the frequency multiplier is 50\,mW. The LO distribution network provides 10\,dB power gain from the reference signal to the LO signals, i.e., $P_{\rm LO,TX}/P_{\rm REF} = P_{\rm LO,RX}/P_{\rm REF} = 10$ in (\ref{LO_TX_Gain}) and (\ref{LO_RX_Gain}). The losses of the LO paths and the power dividers are $L_P = 2\,{\rm dB}$ and $L_D = {\rm 1\, dB}$. The power consumption of the LO distribution network is shown for two values of the power gain efficiency. 

Fig. \ref{fig:PDC_TX_RX_LO} indicates that the power consumption of the LO distribution network is dominant when the number of array elements is small. In this condition, each LO amplifier should provide a high gain to meet the required gain from the reference signal to the LO signals. By increasing the number of array elements, two \textit{competing effects} should be considered. First, the required gain of each LO amplifier is lower, as more LO amplifiers are cascaded and higher gain is achieved. This effect tends to reduce the power consumption of the LO amplifiers. Second, as the number of LO amplifiers increases with the number of array elements, (\ref{LO_TX_amplifiers}) and (\ref{LO_RX_amplifiers}), this effect tends to increase the power consumption. These two competing trends initially reduce the LO power consumption, leading to a flat area shown in Fig. \ref{fig:PDC_TX_RX_LO}, and then increase the LO power consumption as the second effect dominates. 

The power consumption of the TX and RX systems increases with the number of array elements (assuming $N_{\rm TX} = N_{\rm RX}$). For large number of array elements, the power consumption is dominated by the TX power consumption when the power gain efficiency is high, and by the LO distribution network if it is low.

\section{MIMO Radar Back-End}

\subsection{MUSIC and MVDR DOA Estimation Algorithms}

It is assumed that the received signal vector $\mathbf{x}(t)$ comprises $K$ source signals $\mathbf{s}(t)$ with unknown directions and additive white Gaussian noise (AWGN) $\mathbf{n}(t)$. The received signal vector is related to the source signal vector as
\begin{equation}
    \label{x_music}
     \mathbf{x}(t) = \mathbf{A} \mathbf{s}(t) + \mathbf{n}(t).
\end{equation}
The array matrix $\mathbf{A}$ is defined as
\begin{equation}
    \label{A_music}
     \mathbf{A} = [\mathbf{a}(\theta_1, \phi_1), ... ,\mathbf{a}(\theta_K, \phi_K)], 
\end{equation}
where $\mathbf{a}(\theta, \phi) = [a_{mn}(\theta, \phi)]^T$, $a_{mn}(\theta, \phi)$ is the element of array factor in (\ref{AF}), $(\theta_k, \phi_k)$ denotes angular coordinates of the k-th target, and $K$ is the number of targets. 
The covariance matrix of the received signal is estimated using its time-domain samples (snapshots)
\begin{equation}
    \label{Rxx_approx_music}
     \mathbf{\hat{R}}_{xx} = \frac{1}{N_s} \sum _{k=1}^{N_s} \mathbf{x}[k] \mathbf{x}^H[k].
\end{equation}
The covariance matrix and the array matrix are used to construct a spatial spectrum function $P(\theta, \phi)$. In the MUSIC algorithm, $P(\theta, \phi)$ is related to the signal and noise subspace of the covariance matrix \cite{music-paper}, while in the MVDR algorithm it is derived from the inverse of the covariance matrix \cite{mvdr-paper}. Finally, a search algorithm is deployed to find peaks of the spatial spectrum function as DOA of the source signals.

\subsection{DOA Estimation Results}

The MUSIC and MVDR algorithms are evaluated for various architectures of the 2D MIMO radar. The reference distance is assumed as $d_0 = {\rm {1\,m}}$. Root-mean-square error (RMSE) is used as the accuracy metric
\begin{equation}
    \label{rmse}
     {\rm {RMSE}} =   \sqrt{ \frac{1}{KM}  \sum_{j=1}^{M} \sum_{i=1}^{K}\left( (\hat{\phi}_{i,j} - \phi_{i})^2 +  (\hat{\theta}_{i,j} - \theta_{i})^2 \right)},
\end{equation}
where $(\phi_{i}, \theta_{i})$ is the actual DOA of the i-th target, $(\hat{\phi}_{i,j}, \hat{\theta}_{i,j})$ is the estimated DOA of the i-th target in the j-th sample of the Monte Carlo runs, $K$ is the number of targets, and $M$ is the number of the Monte Carlo runs. The noise present in the received signal is modeled as a random Gaussian variable with zero mean and a variance determined based on the SNR, i.e., $n \sim \mathcal{N}(0, \sigma^2)$. We evaluated the simulation results for different numbers of runs and selected $M=100$ to achieve high accuracy with practical simulation times. In this work, it is assumed that there are three targets, $K= 3$, with equal signal amplitudes.

\subsubsection{64-Element Array}
\begin{figure}[!t]
    \centering
    \includegraphics[width = 0.8\columnwidth]{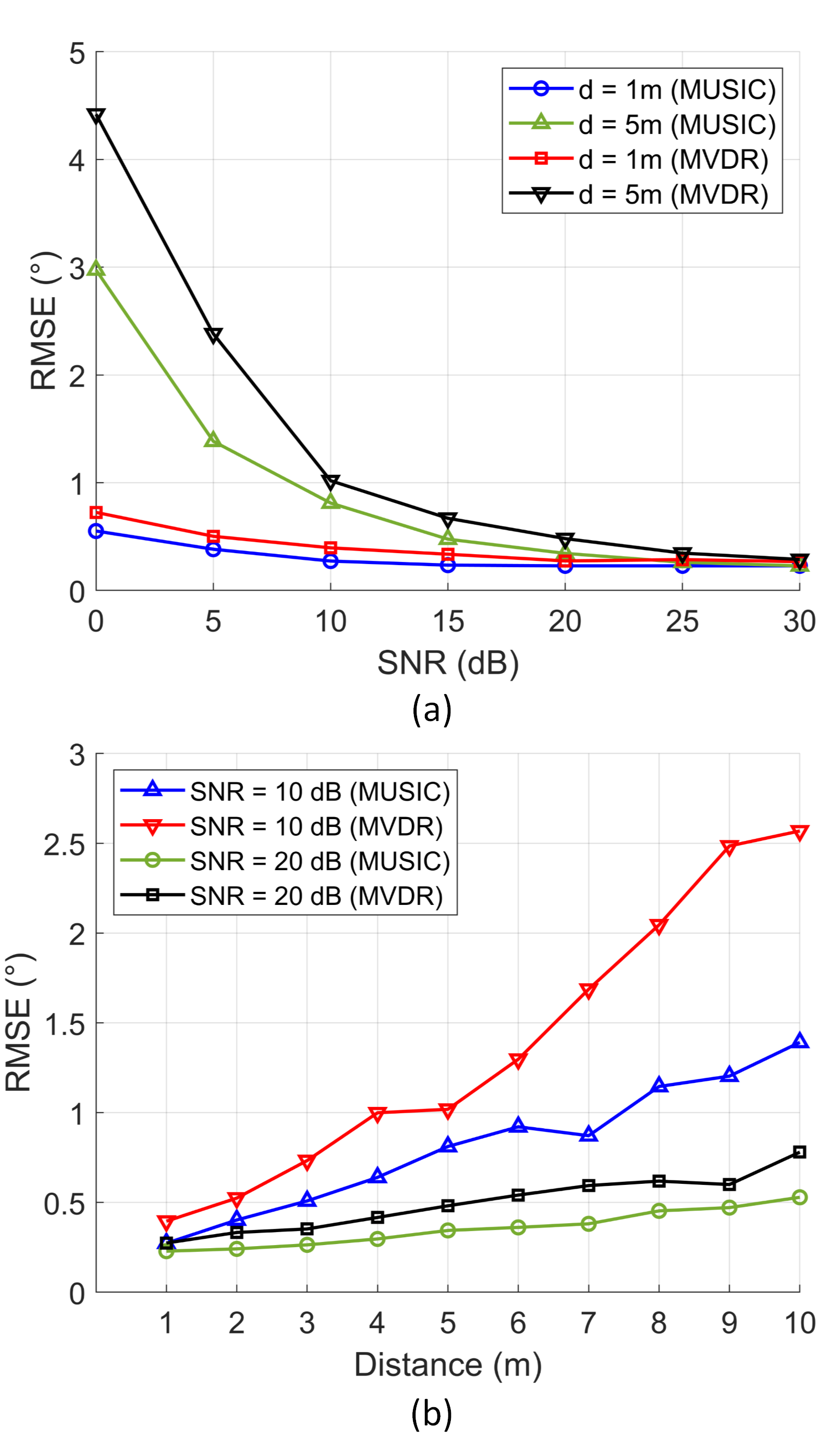}
    \caption{RMSE for the MUSIC and MVDR algorithms with $8\times8$ MIMO radar. For this large array, MUSIC outperforms MVSR especially in the low-SNR conditions.}
    \label{fig:RMSE_8x8}
\end{figure}

For a 64-element array realized as an $8\times8$ square array, the RMSE in terms of SNR and distance is shown in Fig. \ref{fig:RMSE_8x8}. The SNR and the distance are, respectively, varied in the range of 0--30\,dB and 1--10\,m, as typical values in indoor wireless sensing scenarios. The MUSIC algorithm outperforms the MVDR algorithm in the all SNR levels and distances, especially at long-distance and low-SNR conditions. Fig. \ref{fig:RMSE_8x8}(a) indicates that at short distances, e.g., 1\,m, ${\rm RMSE} < 1^{\circ}$ can be maintained even at very low SNR levels. This is the result of an improved SNR of the MIMO array that for the 64-element array is derived using (\ref{snr_mimo}) as $\rm 10\log_{10}(64) \approx 18\,dB$. As shown in Fig. \ref{fig:RMSE_8x8}(b), for the SNR of 20\,dB, ${\rm RMSE} < 0.6^{\circ}$ can be achieved across the 1--10\,m distances using the MUSIC algorithm. 

These results highlight that with a proper algorithm, the required SNR to achieve a specific DOA accuracy (e.g., ${\rm RMSE} < 1^{\circ}$) can be greatly reduced. This leads to much lower required TX output power and, as a result, saves power consumption. The scaling of the TX output power and the RX SNR with distance provide a balance in resource allocation to the short and long distances, and further lowers the power consumption.

\subsubsection{16-Element Array}
\begin{figure}[!t]
    \centering
    \includegraphics[width = 0.8\columnwidth]{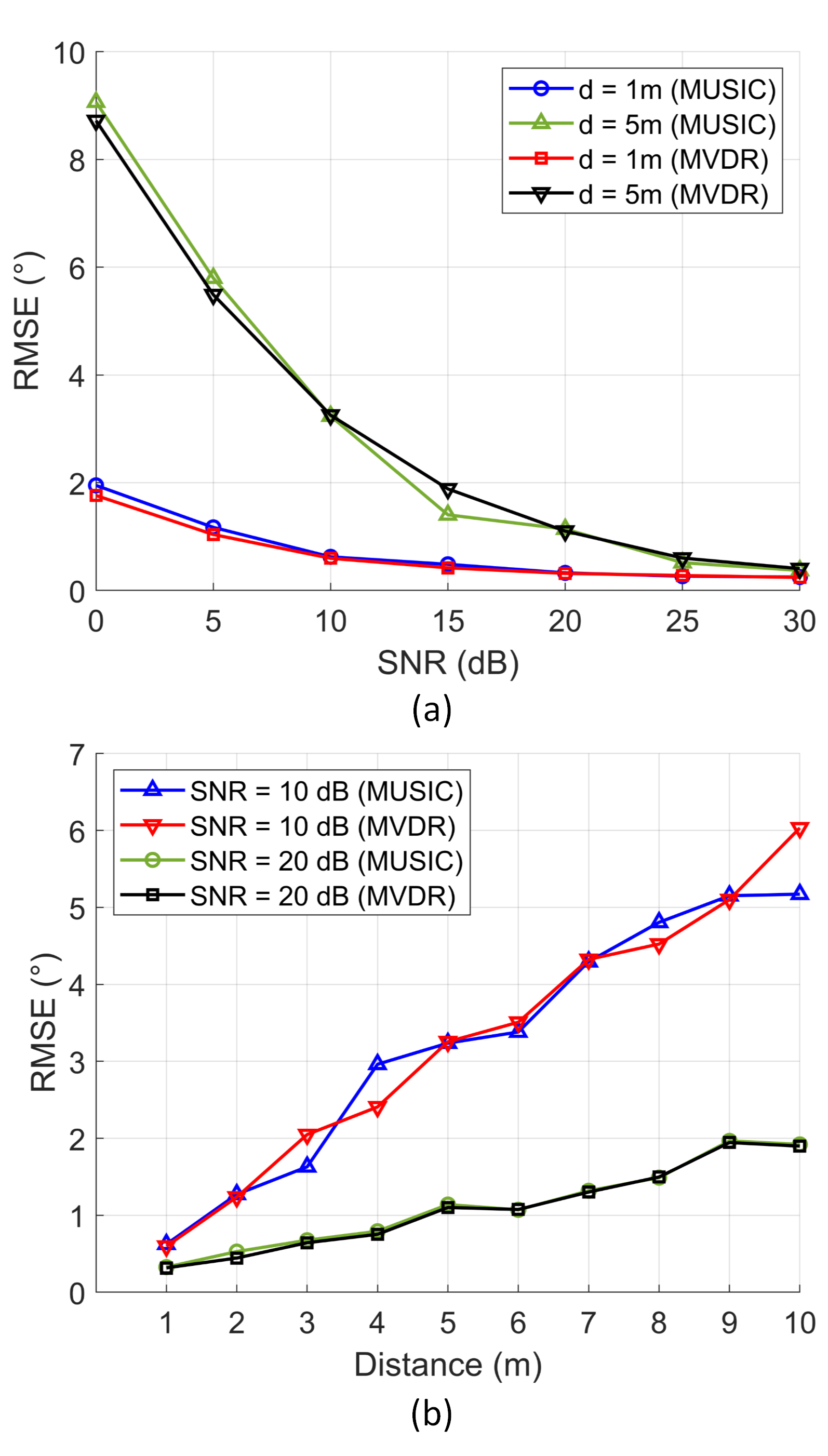}
    \caption{RMSE for the MUSIC and MVDR algorithms with $4\times4$ MIMO radar. In this case, MUSIC and MVDR provide very close results.}
    \label{fig:RMSE_4x4}
\end{figure}

For a 16-element array realized as a $4\times4$ structure, the RMSE in terms of the SNR and distance is shown in Fig. \ref{fig:RMSE_4x4}. Interestingly, the MUSIC and MVDR algorithms provide very close accuracy across the ranges of SNR and distance considered in the simulations. This is in contrary with the $8\times8$ array in which the MUSIC led to higher accuracy compared to the MVDR. The SNR improvement by the MIMO array is $\rm 10\log_{10}(16) \approx 12\,dB$. 
Fig. \ref{fig:RMSE_4x4}(b) indicates that with the 20\,dB SNR, ${\rm RMSE} < 2^{\circ}$ can be achieved in the 1--10\,m range.

\subsubsection{Accuracy of Adaptive Radar}
\begin{figure}[!t]
    \centering
    \includegraphics[width = 0.8\columnwidth]{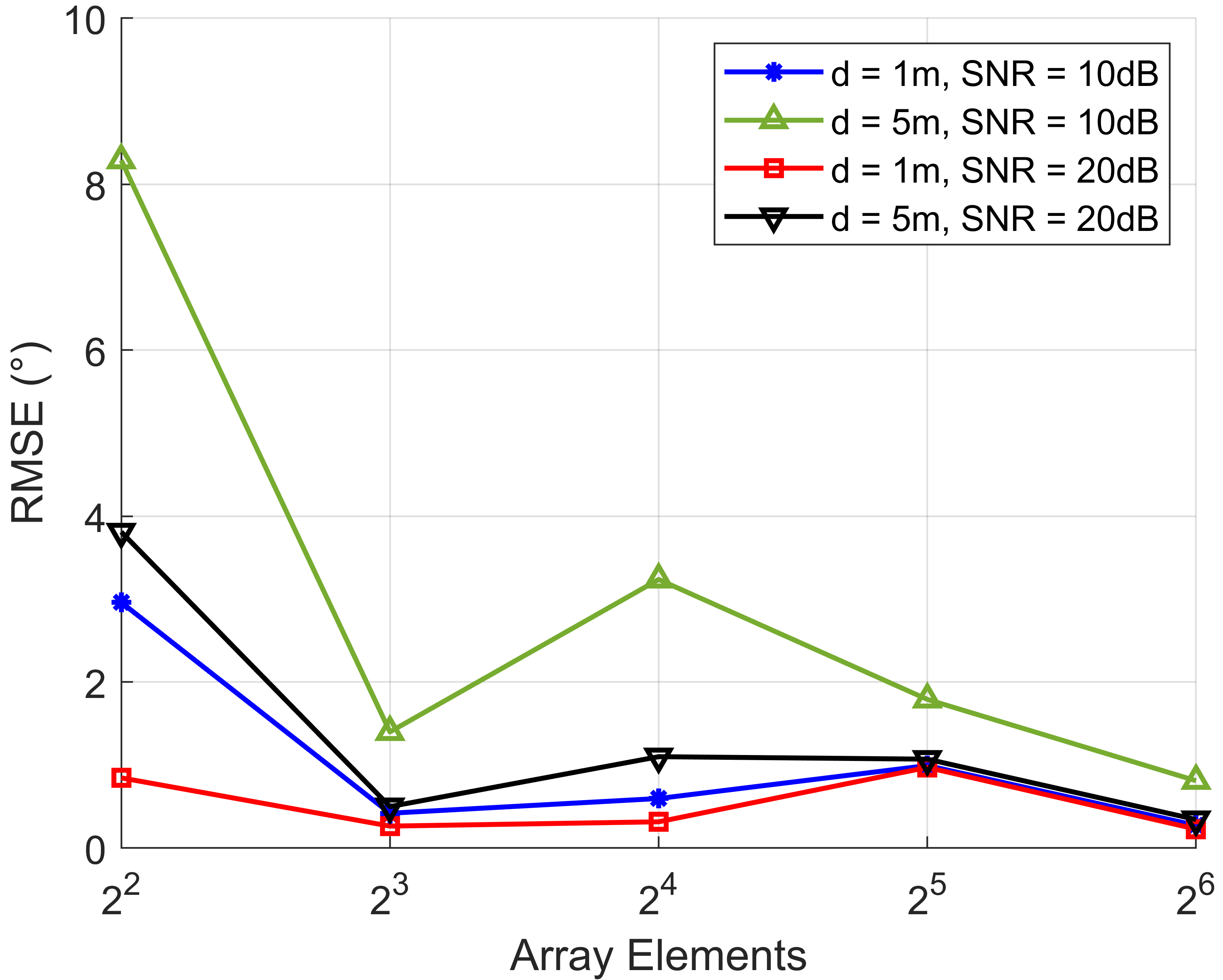}
    \caption{RMSE versus the number of radar array elements. The minimum RMSE achieved using the MUSIC and MVDR algorithms is shown, i.e., $\rm {RMSE} = \min (RMSE_{MUSIC}, RMSE_{MVDR})$.}
    \label{fig:RMSE_elements}
\end{figure}

In Fig. \ref{fig:RMSE_elements}, the RMSE is shown versus the number of array elements. The minimum RMSE achieved using the MUSIC and MVDR algorithms, $\rm {RMSE} = \min (RMSE_{MUSIC}, RMSE_{MVDR})$, is presented. It is noted that a lower RMSE can be achieved for higher SNR and shorter target distance. Increasing the number of array elements generally improves the accuracy, while the amount of improvement varies with the number of array elements, SNR, and distance. The adaptive radar architecture enables trading the accuracy with the array size to improve energy efficiency of the system. For example, when the target moves closer to the radar sensor, the desired accuracy can still be achieved using an smaller array and it is imperative to reconfigure the array structure and reduce power consumption. 

\subsection{Computational Efficiency}

The power consumption of the MIMO radar back-end is dependent on the number of array elements, the size of collected data, the computational complexity of the algorithms, and the features of the digital signal processor (DSP). As most of these information are unknown, we use the \textit{relative simulation time} as a metric for computational efficiency of the radar back-end. Simulations are performed using an Intel Core i5-7300HQ 2.50\,GHz CPU with 8\,GB RAM. The simulation time comprises the time required to complete the DOA estimation simulations for SNR of 0--30\,dB with a step of 5\,dB, distance of 1--10\,m with a step of 1\,m, and 100 Monte Carlo runs. 

\begin{figure}[!t]
   \centering
    \includegraphics[width = \columnwidth]{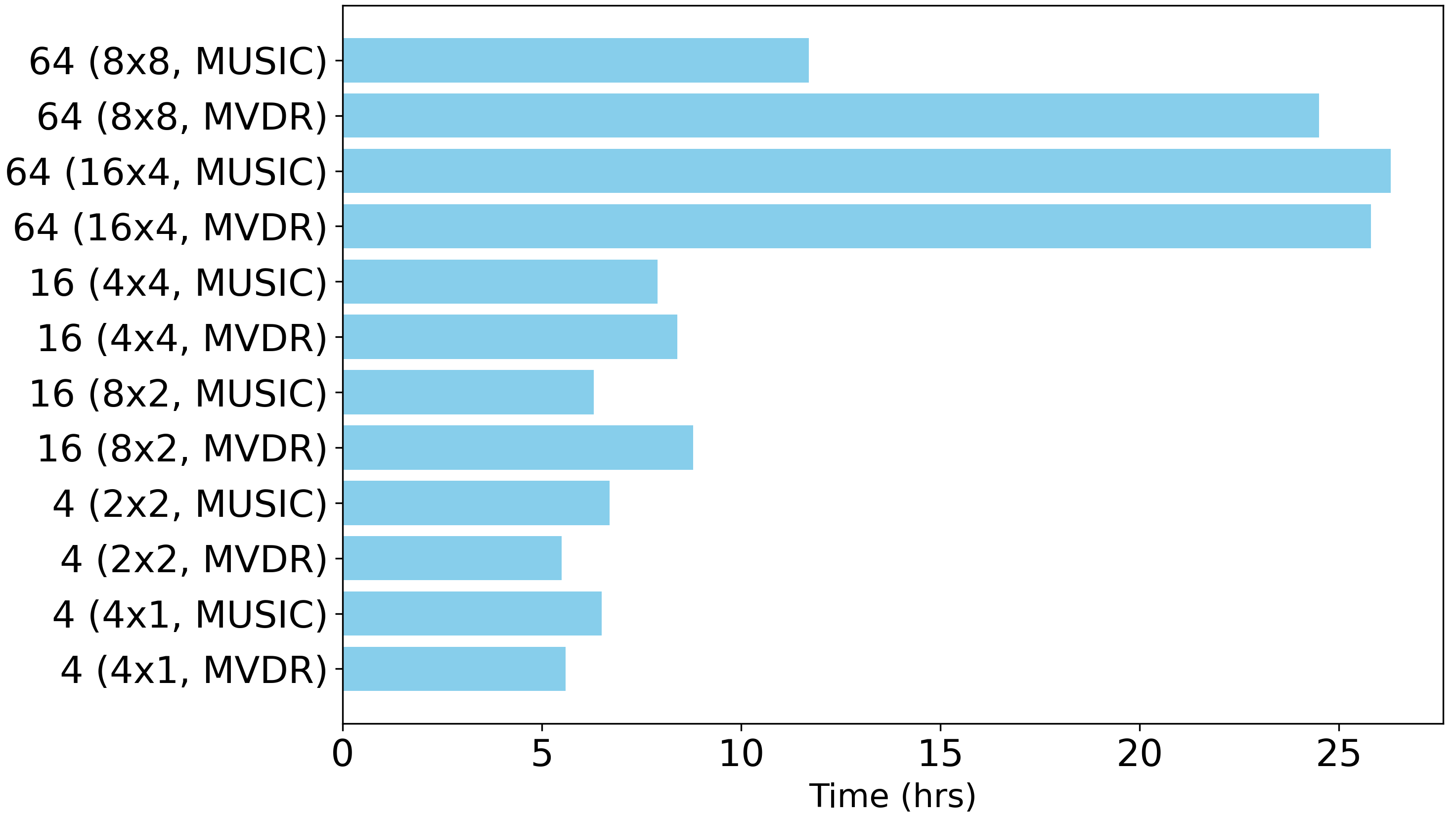}
    \caption{Simulation times of the DOA estimation for various radar array sizes and algorithms.}
   \label{fig:simulation_times}
\end{figure}

Simulation times for various radar array sizes and DOA estimation algorithms are shown in Fig. \ref{fig:simulation_times}. For an $8\times8$ array, the MUSIC algorithm provides much faster simulations, by a factor of about 2x, than the MVDR algorithm. The accuracy is also much higher for the MUSIC algorithm (see Fig. \ref{fig:RMSE_8x8}). Therefore, for an $8\times8$ array the MUSIC algorithm provides the highest performance and energy efficiency. On the other hand, for a $2\times2$ array, the MVDR algorithm outperforms the MUSIC algorithm in terms of the simulation time and also the accuracy. It can be concluded that for large arrays, e.g., over 16 elements, the MUSIC algorithm provides the highest performance while for small arrays, e.g., less than 16 elements, the MVDR algorithm prevails. The adaptive radar architecture allows dynamic reconfiguration of the array size and the selection of the DOA estimation algorithm.

\begin{figure*}[!t]
    \centering
    \includegraphics[width = 1.6 \columnwidth]{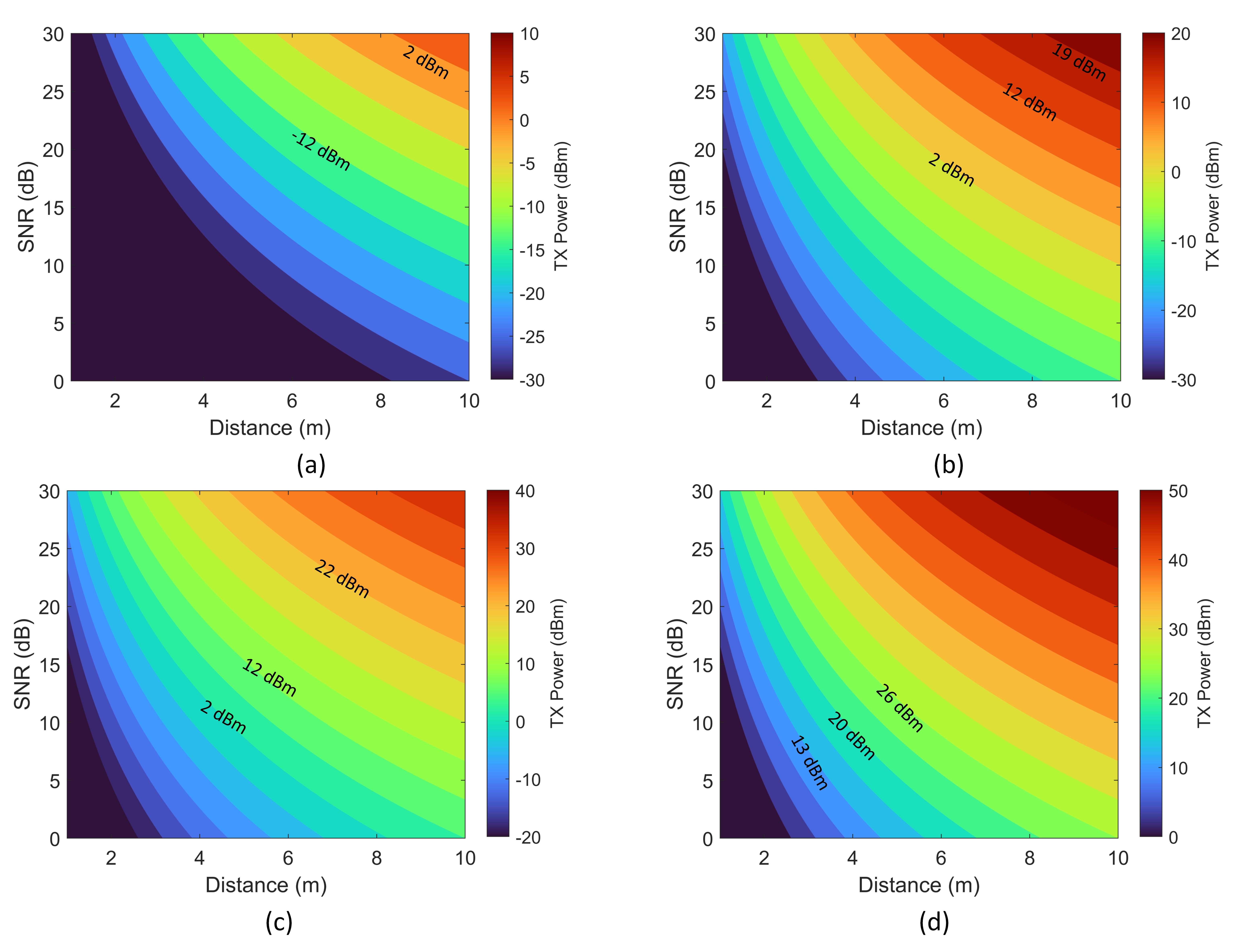}
    \caption{Required TX output power (per radar element) versus the SNR and target distance in various wireless sensing scenarios at the D-band (140\,GHz): (a) free space, (b) clear glass, (c) drywall, (d) wood door.}
    \label{fig:PTX_SNR_distance}
\end{figure*}

\section{Performance of Adaptive MIMO Radar for D-Band Wireless Sensing}

We evaluate performance of the adaptive MIMO radar for D-band wireless sensing and compare it with a conventional MIMO radar. We consider the free-space and through-wall indoor wireless sensing scenarios. It is assumed that there is a dominant line-of-sight (LOS) wave propagation path between the MIMO radar and the target, and the multipath effects are negligible. In the mm-wave bands above 100\,GHz, the multipath signals are severely attenuated by multiple reflections from the indoor objects and walls, and, as a result, their effect is less important than in the lower frequencies \cite{Rappaport-access2019}.

In the following simulations, the radar parameters are assumed as $G_{TX} = G_{RX} = \rm {10\,dB}$, $\lambda = \rm {2.1\,mm}$ (at 140\,GHz), $\sigma = \rm {100\,cm^2}$ as an estimate of the human hand RCS for gesture recognition applications \cite{hugler-gemic2016}, the chirp duration $T_{c} = \rm {10\,\mu s}$, and $\rm {NF_{RX} = 10\,dB}$. The radar frame comprises 100 chirps and the measurement time is $T_{meas} = 100 T_c = {\rm {1\,ms}}$. The TX output power and the RX SNR are scaled with the distance based on a scaling exponent of $p=2$.

\begin{figure}[!t]
    \centering
    \includegraphics[width = 0.8 \columnwidth]{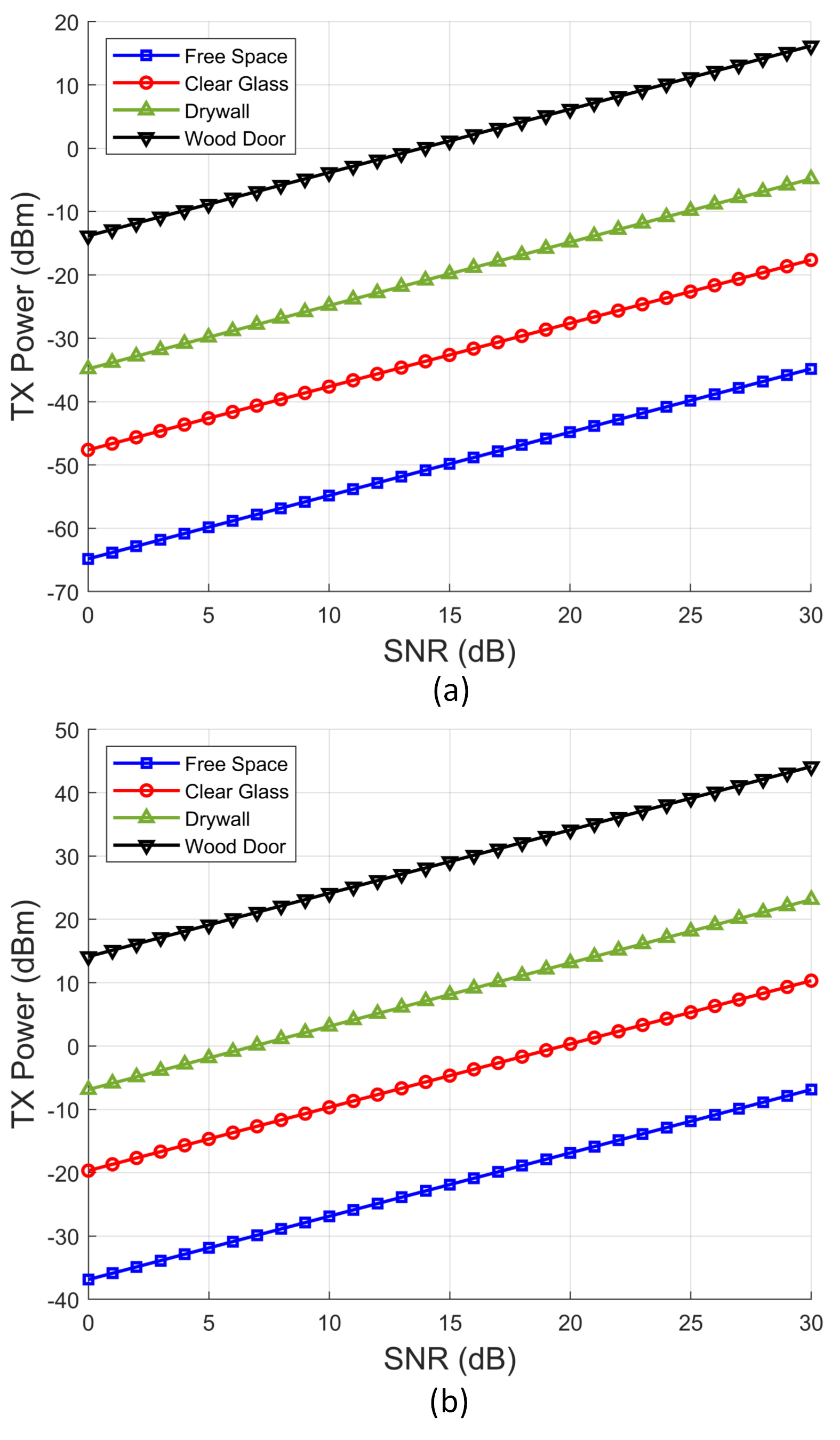}
    \caption{The required TX output power versus the RX SNR for different indoor materials. (a) 1\,m distance, (b) 5\,m distance.}
    \label{fig:PTX_SNR_comparison}
\end{figure}

\begin{figure}[!t]
    \centering
    \includegraphics[width = 0.8 \columnwidth]{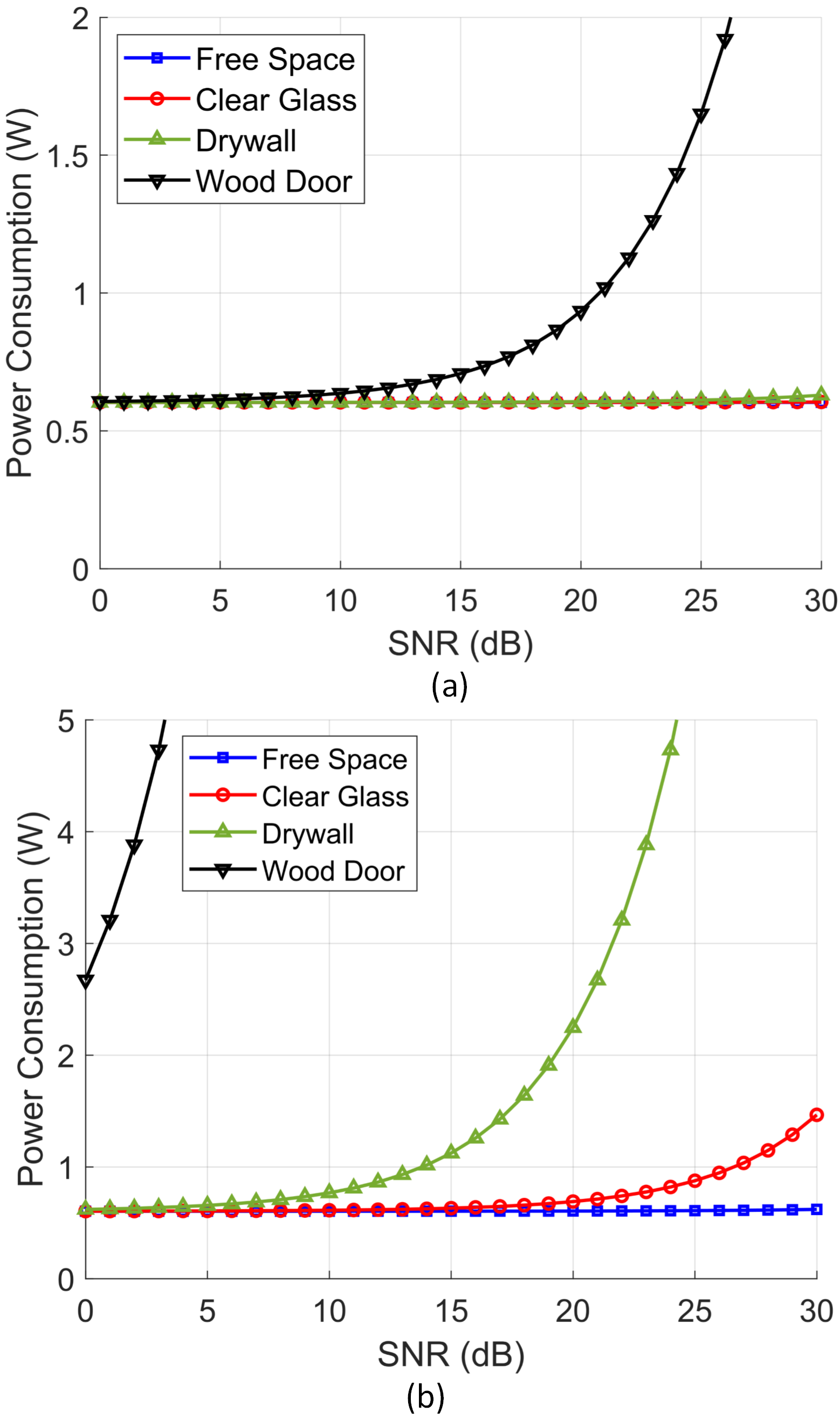}
    \caption{Power consumption of the MIMO radar with 64 elements versus the RX SNR for different indoor materials. (a) 1\,m distance, (b) 5\,m distance.}
    \label{fig:PDC_SNR_comparison}
\end{figure}

\subsection{Free-Space Sensing}
In the free-space wireless sensing, the required TX output power versus the SNR and distance is shown in Fig. \ref{fig:PTX_SNR_distance}(a). The required TX output power can be read less than 0\,dBm in most of the SNR and distance ranges. It reaches to a maximum of approximately 3\,dBm at 30\,dB SNR and 10\,m distance. The required TX output power in the free-space sensing is within the practical power levels of the available D-band radars. Furthermore, if we increase the SNR to 40\,dB to improve the accuracy of the DOA estimation, or for a target with 10x smaller RCS, the required TX output power will increase to 13\,dBm at the extreme distance of 10\,m, which is still within the practical limits. It is important to note that these levels of required TX output power are achieved using our proposed scaling approach with distance. This investigation indicates the promise of high-resolution indoor wireless sensing in the free-space scenario.

\subsection{Through-Wall Sensing}
The through-wall wireless sensing has been successfully realized in the low RF bands (e.g., sub-6\,GHz) \cite{adib-sigcomm2013}. In these bands, the wireless signals can pass through the indoor walls with a moderate loss, but the signal wavelength is too large ($\rm{\lambda > 3\,cm}$) and leads to coarse resolutions. The D-band can provide very fine resolutions ($\rm{\lambda \approx 2\,mm}$), but its feasibility has not been explored in the literature yet. This inspires to use the developed model to investigate the required TX output power and the power consumption of the MIMO radar for the through-wall wireless sensing in the D-band. 

The loss of three representative indoor materials measured at 140\,GHz \cite{Rappaport-access2019} is presented in Table \ref{material_loss}. We assume that the penetration loss through the material is dominant and neglect other losses, including the surface scattering and multipath propagation. The loss of wall medium should be applied twice in the radar SNR to account for the transmitted signal from the radar and the reflected signal from the target
\begin{equation}
    \label{snr_through_wall}
    {\rm {SNR_{\rm TW}}} = \frac{\rm {SNR}}{L_{\rm {wall}}^2},
\end{equation}
where $L_{\rm {wall}}$ is the loss of the wall material.

In Fig. \ref{fig:PTX_SNR_distance}(b), the required TX output power of each element versus the SNR and distance for sensing through clear glass is shown. The glass has 0.6\,cm thickness and 8.6\,dB measured loss at 140\,GHz. The simulation conditions are the same as in the free-space case. A higher TX output power is required to compensate for the additional loss of the wall material. The required TX output power is within the practical limits for most of the SNR and distance ranges, except for extreme values at the top right of the figure. 
In Fig. \ref{fig:PTX_SNR_distance}(c), simulations results are presented for sensing through drywall with 14.5\,cm thickness and 15.0\,dB loss at 140\,GHz. The required TX output power is higher than in the previous case due to the greater loss of drywall. 
In Fig. \ref{fig:PTX_SNR_distance}(d), simulation results are shown for sensing through wood door with 3.5\,cm thickness and 25.5\,dB loss at 140\,GHz. The required TX output power is beyond the practical limits, except for in a small area around the bottom-left of the figure. 

\begin{table}[!t]\footnotesize
\renewcommand{\arraystretch}{1.25}
\centering
\caption{Loss of indoor materials measured at 140\,GHz \cite{Rappaport-access2019}.}
\small\addtolength{\tabcolsep}{12pt}{
\begin{tabular}{c|c|c}
\hline
\textbf{Material} & \textbf{Thickness} & \textbf{Loss}  \\
\hline\hline
Clear Glass & 0.6\,cm & 8.6\,dB  \\
Drywall & 14.5\,cm & 15.0\,dB  \\
Wood Door & 3.5\,cm & 25.5\,dB  \\
\hline
\end{tabular}%
}
\label{material_loss}
\end{table}


A comparison of the required TX output power versus the RX SNR for different indoor materials is shown in Fig. \ref{fig:PTX_SNR_comparison}. The results are shown at two distances 1\,m and 5\,m. The required TX output power at the distance of 1\,m is very low in the all conditions. This is a promising result for the future of large MIMO radars in short indoor distances. However, the required TX output power at 5\,m distance can become very high and impractical in some conditions. 

The power consumption of the MIMO radar with 64 elements versus the RX SNR is shown in Fig. \ref{fig:PDC_SNR_comparison}. In the low SNR, the requited TX power is low and the power consumption of the array is dominated by that of RX and LO, which are independent of the SNR. The required TX power becomes significant for the high SNR and large loss of indoor material, e.g., the wood door. This results in a power consumption sharply increasing with the SNR. The power consumption is higher for a longer target distance.


\section{Conclusion}

In this paper, we proposed an adaptive MIMO radar architecture for energy-efficient wireless sensing. The effective array size is reconfigured based on the required radar resolution, while the receiver SNR and the transmitter output power are scaled with the distance. Additionally, the direction-finding algorithm can be dynamically selected to improve the accuracy and reduce the computational complexity. We evaluated performance of the adaptive MIMO radar for wireless sensing in the D-band and presented the advantages over a conventional MIMO radar.

\section*{REFERENCES}

\def\refname{}


\begin{thebibliography}{34}\vspace*{-12pt}

\bibitem{6g-white-2020}
C. de Lima \textit{et al.}, (Eds.). (2020). 6G white paper on localization and sensing. 6G Research Visions, No. 12. University of Oulu. http://urn.fi/ urn:isbn:9789526226743

\bibitem{Rappaport-access2019}
T. S. Rappaport \textit{et al.}, ``Wireless communications and applications above 100 GHz: Opportunities and challenges for 6G and beyond," \textit{IEEE Access}, vol. 7, pp. 78729--78757, June 2019, doi: 10.1109/ACCESS.2019.2921522.

\bibitem{maiwald-proc2022}
T. Maiwald \textit{et al.}, ``A review of integrated systems and components for 6G wireless communication in the D-band," \textit{Proc. IEEE}, vol. 111, no. 3, pp. 220--256, March 2023, doi: 10.1109/JPROC.2023.3240127.

\bibitem{kim-tap2015}
S. Kim, W. T. Khan, A. Zajic and J. Papapolymerou, ``D-band channel measurements and characterization for indoor applications," \textit{IEEE Trans. Antennas Propag.}, vol. 63, no. 7, pp. 3198--3207, July 2015, doi: 10.1109/TAP.2015.2426831.

\bibitem{beelde-access2021}
B. De Beelde, D. Plets, C. Desset, E. Tanghe, A. Bourdoux and W. Joseph, ``Material characterization and radio channel modeling at D-band frequencies," \textit{IEEE Access}, vol. 9, pp. 153528--153539, Nov. 2021, doi: 10.1109/ACCESS.2021.3127399.




\bibitem{Visweswaran-jssc2021}
A. Visweswaran \textit{et al.}, ``A 28-nm-CMOS based 145-GHz FMCW radar: System, circuits, and characterization," \textit{IEEE J. Solid-State Circuits}, vol. 56, no. 7, pp. 1975--1993, July 2021, doi: 10.1109/JSSC.2020.3041153.

\bibitem{sense-lmwc-2022}
B. Sene, D. Reiter, H. Knapp, H. Li, T. Braun and N. Pohl, ``An automotive D-band FMCW radar sensor based on a SiGe-transceiver MMIC," \textit{IEEE
Microw. Wireless Compon. Lett}, vol. 32, no. 3, pp. 194--197, March 2022, doi: 10.1109/LMWC.2021.3121656.

\bibitem{Furqan-fullyIntgrated-2019}
M. Furqan, F. Ahmed and A. Stelzer, ``A D-band fully-integrated 2-RX, 1-TX FMCW radar sensor with 13dBm output power,"  \textit{Eur. Microwave Integrated Circuits Conf. (EuMIC)}, Paris, France, 2019, pp. 100--103, doi: 10.23919/EuMIC.2019.8909441.

\bibitem{Ahmad-W-Band-D-Band}
W. A. Ahmad \textit{et al.}, ``Multimode W-band and D-band MIMO scalable radar platform," \textit{IEEE Trans. Microw. Theory Techn.}, vol. 69, no. 1, pp. 1036--1047, Jan. 2021, doi: 10.1109/TMTT.2020.3038532.

\bibitem{Deng-D-Band-2023}
W. Deng \textit{et al.}, ``A D-band joint radar-communication CMOS transceiver," \textit{IEEE J. Solid-State Circuits}, vol. 58, no. 2, pp. 411--427, Feb. 2023, doi: 10.1109/JSSC.2022.3185160.

\bibitem{Zandieh-155GHz-FMCW}
A. Zandieh, S. Bonen, M. S. Dadash, M. J. Gong, J. Hasch and S. P. Voinigescu, ``155 GHz FMCW and stepped-frequency carrier OFDM radar sensor transceiver IC featuring a PLL with $<$30 ns settling time and 40 fs rms jitter," \textit{IEEE Trans. Microw. Theory Techn.}, vol. 69, no. 11, pp. 4908--4924, Nov. 2021, doi: 10.1109/TMTT.2021.3094189.

\bibitem{Kucharski-Multi-Purpose-2017}
H. J. Ng, M. Kucharski, W. Ahmad and D. Kissinger, ``Multi-purpose fully differential 61- and 122-GHz radar transceivers for scalable MIMO sensor platforms," \textit{IEEE J. Solid-State Circuits}, vol. 52, no. 9, pp. 2242--2255, Sept. 2017, doi: 10.1109/JSSC.2017.2704602.

\bibitem{D-Band-Low-Power-2022}
S. Park \textit{et al.}, ``A D-band low-power and high-efficiency frequency multiply-by-9 FMCW radar transmitter in 28-nm CMOS," \textit{IEEE J. Solid-State Circuits}, vol. 57, no. 7, pp. 2114--2129, July 2022, doi: 10.1109/JSSC.2022.3157643.

\bibitem{nikandish-jmw2024}
R. Nikandish, ``GaN system-on-chip: Pushing the limits of integration and functionality," \textit{IEEE J. Microwaves}, 2024, doi: 10.1109/JMW.2024.3429615.


\bibitem{fishler-radar2004}
E. Fishler, A. Haimovich, R. Blum, D. Chizhik, L. Cimini and R. Valenzuela, ``MIMO radar: an idea whose time has come," \textit{Proc.
IEEE Radar Conf.}, 2004, pp. 71--78, doi: 10.1109/NRC.2004.1316398.

\bibitem{li-2007}
J. Li and P. Stoica, ``MIMO radar with colocated antennas," \textit{IEEE Signal Process. Mag.}, vol. 24, no. 5, pp. 106--114, Sept. 2007, doi: 10.1109/MSP.2007.904812.

\bibitem{serio-2020}
A. Di Serio, P. Hügler, F. Roos and C. Waldschmidt, ``2-D MIMO radar: A method for array performance assessment and design of a planar antenna array," \textit{IEEE Trans. Antennas Propag.}, vol. 68, no. 6, pp. 4604--4616, June 2020, doi: 10.1109/TAP.2020.2972643.

\bibitem{zhuge-tap2012}
X. Zhuge and A. G. Yarovoy, ``Study on two-dimensional sparse MIMO
UWB arrays for high resolution near-field imaging," \textit{IEEE Trans. Antennas Propag.}, vol. 60, no. 9, pp. 4173--4182, Sep. 2012.

\bibitem{peng-tmtt2018}
Z. Peng and C. Li, ``A portable K-band 3-D MIMO radar with nonuniformly spaced array for short-range localization," \textit{IEEE Trans. Microw. Theory Techn.}, vol. 66, no. 11, pp. 5075--5086, Nov. 2018, doi: 10.1109/TMTT.2018.2869565.

\bibitem{tan-tap2017}
K. Tan \textit{et al.}, ``On sparse MIMO planar array topology optimization for UWB near-field high-resolution imaging," \textit{IEEE Trans. Antennas Propag.}, vol. 65, no. 2, pp. 989--994, Feb. 2017, doi: 10.1109/TAP.2016.2632626.

\bibitem{Puglielli-proc2016}
A. Puglielli \textit{et al.}, ``Design of energy- and cost-efficient massive MIMO arrays," \textit{Proc. IEEE}, vol. 104, no. 3, pp. 586--606, March 2016, doi: 10.1109/JPROC.2015.2492539.

\bibitem{Torkildson-twc2011}
E. Torkildson, U. Madhow and M. Rodwell, ``Indoor millimeter wave MIMO: Feasibility and performance," \textit{IEEE Trans. Wireless
Commun.}, vol. 10, no. 12, pp. 4150--4160, December 2011, doi: 10.1109/TWC.2011.092911.101843.

\bibitem{nikandish-tradar2023}
R. Nikandish, A. Yousefi and E. Mohammadi, ``Spurs in millimeter-wave FMCW radar system-on-chip," \textit{IEEE Trans. Radar Syst.}, vol. 1, pp. 21--33, Dec. 2023, doi: 10.1109/TRS.2023.3265845.

\bibitem{arnold-tap2019}
B. T. Arnold and M. A. Jensen, ``The effect of antenna mutual coupling on MIMO radar system performance," \textit{IEEE Trans. Antennas Propag.}, vol. 67, no. 3, pp. 1410--1416, March 2019, doi: 10.1109/TAP.2018.2888702.

\bibitem{music-paper}
R. Schmidt, ``Multiple emitter location and signal parameter estimation," \textit{IEEE Trans. Antennas Propag.}, vol. 34, no. 3, pp. 276--280, March 1986, doi: 10.1109/TAP.1986.1143830.

\bibitem{mvdr-paper}
J. Capon, ``High–resolution frequency-wavenumber spectrum analysis,” \textit{Proc. IEEE}, 1969, Vol. 57, pp. 1408--1418

\bibitem{sun-spm-2020}
S. Sun, A. P. Petropulu and H. V. Poor, ``MIMO radar for advanced driver-assistance systems and autonomous driving: Advantages and challenges," \textit{IEEE Signal Process. Mag.}, vol. 37, no. 4, pp. 98--117, July 2020, doi: 10.1109/MSP.2020.2978507.


\bibitem{Stutzman-2012}
W. L. Stutzman and G. A. Thiele. \emph{Antenna Theory and Design.} Third Ed., John Wiley \& Sons, 2012.

\bibitem{gu-jmw2021}
X. Gu, D. Liu and B. Sadhu, ``Packaging and antenna integration for silicon-based millimeter-wave phased arrays: 5G and beyond," \textit{IEEE J. Microwaves}, vol. 1, no. 1, pp. 123-134, Jan. 2021, doi: 10.1109/JMW.2020.3032891.

\bibitem{tang-tcas2020}
X. Tang \textit{et al.}, ``Design of D-band transformer-based gain-boosting class-AB power amplifiers in silicon technologies," \textit{IEEE Trans. Circuits Syst. I: Reg. Papers}, vol. 67, no. 5, pp. 1447-1458, May 2020, doi: 10.1109/TCSI.2020.2974197.

\bibitem{Khatibi-tmtt2018}
H. Khatibi, S. Khiyabani, and E. Afshari, ``A 173 GHz amplifier with a 18.5 dB power gain in a 130 nm SiGe process: A systematic design of high-gain amplifiers above $f_{max}/2$," \textit{IEEE Trans. Microw. Theory Techn.}, vol. 66, no. 1, pp. 201-214, Jan. 2018, doi: 10.1109/TMTT.2017.2727038.

\bibitem{park-jssc2019}
D.-W. Park, D. R. Utomo, B. H. Lam, S.-G. Lee, and J.-P. Hong, ``A 230–260-GHz wideband and high-gain amplifier in 65-nm CMOS based on dual-peak $G_{max}$-core," \textit{IEEE J. Solid-State Circuits}, vol. 54, no. 6, pp. 1613–1623, Jun. 2019, doi: 10.1109/JSSC.2019.2899515.


\bibitem{hugler-gemic2016}
P. Hügler, M. Geiger and C. Waldschmidt, ``RCS measurements of a human hand for radar-based gesture recognition at E-band," \textit{German Microwave Conference (GeMiC)}, Bochum, Germany, 2016, pp. 259--262, doi: 10.1109/GEMIC.2016.7461605.

\bibitem{adib-sigcomm2013}
F. Adib and D. Katabi, ``See through walls with WiFi!," \textit{SIGCOMM Comput. Commun.} Rev. 43, 4, Oct. 2013, 75--86. https://doi.org/10.1145/2534169.2486039

\bibitem{zhang-isscc2023}
J. Zhang, A. Singhvi, S. S. Ahmed and A. Arbabian, ``18.1 A W-band transceiver array with 2.4 GHz LO synchronization enabling full scalability for FMCW radar," \textit{IEEE Int. Solid-State Circuits Conf. (ISSCC) Dig. Tech. Papers}, San Francisco, CA, USA, 2023, pp. 282--284, doi: 10.1109/ISSCC42615.2023.10067317.

\end{thebibliography}
\end{document}